\pdfoutput=1
\documentclass[a4paper,11pt]{article}
\usepackage{jheppub} % for details on the use of the package, please see the JINST-author-manual
\usepackage{comment}
\usepackage{amsmath,amssymb,amsfonts,mathtools}
\usepackage[toc,page]{appendix}
\usepackage{hyperref}
\usepackage{physics}
\usepackage{float}
\usepackage{slashed}
\usepackage[dvipsnames]{xcolor}
\usepackage{tikz}
\usetikzlibrary{decorations.markings, arrows.meta, decorations.pathmorphing}

\newcommand{\lara}[1]{\left\langle #1 \right\rangle}

\newcommand{\ab}[0]{{\alpha \beta}}

\newcommand{\mn}[0]{{\mu \nu}}

\newcommand{\MN}[0]{{M N}}

\newcommand{\AB}[0]{{AB}}

\renewcommand{\ij}[0]{{ij}}

\newcommand{\pd}[0]{\partial}

\newcommand{\Psib}[0]{\Bar{\Psi}}

\newcommand{\ub}[0]{\Bar{u}}

\newcommand{\mt}[1]{\textrm{\scriptsize #1}}

\newcommand{\A}[0]{\mathcal{A}}
\newcommand{\Ag}[0]{\mathcal{A}_\text{g}}
\newcommand{\Bg}[0]{\mathcal{B}_\text{g}}

\newcommand{\Apol}[0]{\mathcal{A}_\text{pol}}
\newcommand{\B}[0]{\mathcal{B}}

\newcommand{\D}[0]{\mathcal{D}}
\newcommand{\Dg}[0]{\mathcal{D}_\text{g}}

\newcommand{\Dpol}[0]{\mathcal{D}_\text{pol}}

\newcommand{\F}[0]{\mathcal{F}}
\newcommand{\G}[0]{\mathcal{G}}

\newcommand{\K}[0]{\mathcal{K}}
\renewcommand{\L}[0]{\mathcal{L}}

\newcommand{\N}[0]{\mathcal{N}}

\newcommand{\R}[0]{\mathcal{R}}

\newcommand{\V}[0]{\mathcal{V}}

\newcommand{\oford}[1]{\mathcal{O}\qty(#1)}
\newcommand{\liegroup}[3]{\text{#1}\qty(#2)^{#3}}

\renewcommand{\dd}{\mathrm{d}}
\def\Nc{N_\mt{c}}

\def\MP{M_\mt{P}}

\def\mN{m_\mt{N}}
\def\VN{\V_\mt{N}}

\def\VR{\V_{R}}
\def\VL{\V_{L}}
\def\vN{v_{\mt{N}}}

\newcommand{\be}{\begin{equation}}
\newcommand{\ee}{\end{equation}}
\newcommand{\bea}{\begin{eqnarray}}
\newcommand{\eea}{\end{eqnarray}}

\makeatletter
\def\@fpheader{\relax}
\makeatother

\preprint{$\begin{array}{rr}
	\text{HIP-2026-1/TH}\\\text{APCTP Pre2026 - 001}\end{array}$
}

\title{Gravitational form factors of the proton \\ in the improved holographic QCD model}

\author[a]{Antti Hippel\"ainen,}
\author[a]{Niko Jokela,} 
\author[b,c,d]{and Matti J\"arvinen}
\affiliation[a]{Department of Physics and Helsinki Institute of Physics, \\
P.O. Box 64, FI-00014, University of Helsinki, Finland}
\affiliation[b]{Institute of Theoretical Physics, Chinese Academy of Sciences, \\
55 Zhong Guan Cun East Road, Haidian District, Beijing 100190, China}
\affiliation[c]{Asia Pacific Center for Theoretical Physics, \\
77 Cheongam-ro, Nam-gu, Pohang 37673, Republic of Korea}
\affiliation[d]{Department of Physics, Pohang University of Science and Technology, \\
77 Cheongam-ro, Nam-gu, Pohang 37673, Republic of Korea}

\emailAdd{antti.hippelainen@helsinki.fi}
\emailAdd{niko.jokela@helsinki.fi}
\emailAdd{mattijarvinen@itp.ac.cn}

\abstract{We compute the gluonic contribution to the gravitational form factors of the proton using the improved holographic QCD model, in which the proton is described in terms of bulk Dirac fermions.  Model parameters are constrained using lattice and phenomenological input, allowing us to obtain estimates for the gravitational form factors and to compare them with results from other approaches. Using the extracted form factors, we analyze mechanical properties of the proton, including pressure and shear distributions. We obtain estimates for the mechanical and mass radii of the proton, $\rho_{\text{mech}} \approx 0.63$ fm, $\rho_{\text{mass}} \approx  0.65$ fm, and compare these values with radius estimates from other nonperturbative studies.}

\begin{document}
\maketitle
\flushbottom
\setcounter{page}{2}

\clearpage

%%%%%%%%%%%%%%%%%%%%%%%%%%%%%%%%%%%%%%%%%%%%%%%%%%
%%%%%%%%%%%%%%%%%%%%%%%%%%%%%%%%%%%%%%%%%%%%%%%%%%
\section{Introduction}
%%%%%%%%%%%%%%%%%%%%%%%%%%%%%%%%%%%%%%%%%%%%%%%%%%
%%%%%%%%%%%%%%%%%%%%%%%%%%%%%%%%%%%%%%%%%%%%%%%%%%

Probing the structure of the proton in the strongly coupled regime of quantum chromodynamics (QCD) remains a challenging problem. While the proton contains three valence quarks, its full structure involves a complex interplay of gluons and sea quarks that is still not fully understood.

The internal structure of the proton will be explored in unprecedented detail at the upcoming Electron-Ion Collider~\cite{Accardi:2012qut}. For the first time, spin-polarized beams of protons will be used, allowing for precision studies of spin-dependent properties and offering new insights into the long-standing proton spin puzzle~\cite{Ji:2020ena}. Earlier experiments, such as HERA~\cite{Habib:2010zz}, have also contributed significantly to our understanding of proton structure.

The experimental studies of the proton focus on deep inelastic scattering (DIS) experiments. In DIS, a lepton, typically an electron, scatters off a hadron such as a proton, thereby probing its internal constituents~\cite{Collins:1350496}. In inclusive DIS, the hadronic final state is not fully reconstructed. Hard exclusive scattering processes, in which the proton remains intact and all final-state particles are detected, can provide complementary information to DIS and are sensitive to detailed aspects of the internal structure of the proton~\cite{Goeke:2001tz}. 

A key theoretical challenge is to access nonperturbative features of hadronic structure. While lattice QCD provides a first-principles nonperturbative approach, extracting spatially resolved and dynamical hadronic observables remains technically demanding, motivating the exploration of complementary theoretical frameworks. Holographic QCD models, inspired by the string/gravity correspondence, provide such a framework for this purpose. These models allow for a dual gravitational description of strongly coupled gauge theories, offering semi-quantitative predictions for observables that are otherwise difficult to compute. Over the past two decades, holography has been successfully applied to a wide range of problems in QCD-like theories, including hadron spectra, thermodynamics, and transport~\cite{Ramallo:2013bua}.

In this work, we study the gravitational form factors (GFFs) of the proton using a holographic approach. These form factors, $\A,\B,\D$, are defined through matrix elements of the energy-momentum tensor $T^{\mu\nu}(x)$ in a proton state~\cite{Polyakov:2018zvc, Abidin:2009hr},
\be\label{eq:meldecomposition}
\begin{split}
 &\mel{p_2, s_2}{T^\mn(x)}{p_1,s_1} = \\
 &\Bar{u}(p_2, s_2) \qty(\A(t)\frac{P^{\{ \mu}\gamma^{\nu\}}}{2}+ \B(t) \frac{i P^{\{\mu}\sigma^{\nu \}\alpha} k_\alpha}{4\mN} + \D(t) \frac{k^\mu k^\nu - \eta^\mn k^2}{4 \mN})u(p_1, s_1)e^{i k\cdot x} \ , 
\end{split}
\ee
where $p_1$ ($p_2$) and $s_1$ ($s_2$) are the on-shell momentum and spin of the incoming (outgoing) proton. For most of the work, we keep the spin indices implicit. Additionally, $t = k^2 \equiv -K^2$ with $k=p_2-p_1$ being the exchanged momentum, bracketed indices $\mu,\nu,\alpha,\ldots=0,1,2,3$ are symmetrized as $a^{\{\mu}b^{\nu\}}=(a^\mu b^\nu+a^\nu b^\mu)$, $\mN$ is the mass of the proton, $P = (p_1 + p_2)/2$, and $\sigma^\mn = \frac{i}{2}\comm{\gamma^\mu}{\gamma^\nu}$. The sign convention for $\eta$ is mostly minus.

The gravitational form factors encode information about the distribution of energy, momentum, and internal forces inside the proton. {\color{black}These mechanical properties of the proton are not directly accessible experimentally and are currently inferred only through model-dependent analyses of exclusive scattering data~\cite{Burkert:2018bqq,Burkert:2021ith, Duran:2022xag, 007:2026dow, CLAS:2026lls}, through dispersion relations~\cite{Cao:2024zlf,Cao:2025dkv}, or via lattice analysis~\cite{Shanahan:2018nnv,Hackett:2023rif}.} This motivates complementary non-perturbative theoretical approaches that can provide internally consistent predictions for these quantities. To this end, previous holographic studies of GFFs for nucleons and mesons include~\cite{Abidin:2008ku,Abidin:2009hr,Mamo:2019mka,Fujita:2022jus,Li:2023izn,Sugimoto:2025btn,Deng:2025fpq,Liu:2025vfe}; see also~\cite{Hashimoto:2008zw} for an early review of holographic form factors. In particular, the D-term or druck term, defined as the value of $\D(t)$ in the forward limit $t = 0$, remains poorly understood and has recently begun to attract interest also within the holographic community~\cite{Fujita:2022jus,Sugimoto:2025btn}. 

In this work we focus only on the gluonic contribution to the form factors. We work with the improved holographic QCD (IHQCD) model~\cite{Gursoy:2010fj,Gursoy:2007cb,Gursoy:2007er}, a holographic description of pure Yang--Mills theory that provides a phenomenologically realistic background geometry and has been successfully applied to a wide range of hadronic and thermodynamic observables. The background geometry is obtained by solving the five-dimensional gravitational equations of motion derived from the bulk action. The proton is modeled through a pair of bulk Dirac spinors. Matrix elements of the energy-momentum tensor are obtained via the holographic dictionary from the couplings induced by diffeomorphism invariance of the spinor action, allowing us to extract the GFFs of the proton. From these, we compute several quantities related to hadronic structure, including the gluonic contribution to the $\D(t)$ form factor, denoted as $\Dg(t)$, and mechanical properties of the proton.

The free parameters of the model are fixed by fitting the form factor $\Ag(t)$, {\emph{i.e.}}, the gluonic contribution to $\A(t)$, to two sets of lattice data~\cite{Shanahan:2018pib,Hackett:2023rif}, after which the form factor $\Dg(t)$ is determined without further adjustment. Using the extracted form factors, we compute radial pressure and shear distributions inside the proton and compare them to existing theoretical and lattice-based studies. Finally, the proton radii obtained in our model, as determined by the gluonic distributions, are $\rho_{\text{mech}} \approx  0.63$ fm and $\rho_{\text{mass}} \approx  0.65$ fm.

The paper is organized as follows. In Section~\ref{sec:holographicsetting}, we present the holographic model used to describe the background geometry and external sources. In Section~\ref{sec:formfactors}, we compute the relevant couplings and extract the gravitational form factors. In Section~\ref{sec:mechanicalstructure}, we analyze the mechanical structure of the proton implied by these form factors and compare our results to other theoretical approaches. We conclude in Section~\ref{sec:conclusions}. Technical details are provided in Appendices~\ref{app:amplitudescouplingsvertices} and \ref{app:IHQCD}.

%%%%%%%%%%%%%%%%%%%%%%%%%%%%%%%%%%%%%%%%%%%%%%%%%%
%%%%%%%%%%%%%%%%%%%%%%%%%%%%%%%%%%%%%%%%%%%%%%%%%%
\section{Holographic model}\label{sec:holographicsetting}
%%%%%%%%%%%%%%%%%%%%%%%%%%%%%%%%%%%%%%%%%%%%%%%%%%
%%%%%%%%%%%%%%%%%%%%%%%%%%%%%%%%%%%%%%%%%%%%%%%%%%

The five-dimensional gravitational model we work with is IHQCD~\cite{Gursoy:2007cb,Gursoy:2007er}, a string-inspired model for pure glue Yang--Mills theory. Apart from the five-dimensional metric, which, as usual, is dual to the energy-momentum tensor in field theory, the model contains a dynamical dilaton field $\Phi$. This field is interpreted to be dual to the $\mathrm{Tr}\,G_{\mu\nu}G^{\mu\nu}$ operator where $G$ is the field strength tensor of gluons and the trace is over color indices. Therefore, the dictionary of the model includes duals for the most important operators at weak coupling, {\emph{i.e.}}, the ultraviolet (UV) marginal operators of the theory.

The action of IHQCD reads 
\begin{equation}
\label{eq:IHQCDaction}
    S_{\text{IHQCD}} = \MP^3 \Nc^2 \int \dd^5 x \sqrt{-\det g}\, \qty[\R - \frac{4}{3}(\partial \Phi)^2 + V(\Phi)] + 2\MP^3 \Nc^2 \int_\partial \dd^4 x \sqrt{-\det h}\,\K \ ,
\end{equation}
where $\MP$ is the Planck mass, $\Nc$ is the number of colors, $\R$ is the Ricci scalar, $h$ is the induced metric on the four-dimensional boundary manifold, and $\K$ is the extrinsic curvature. The multiplier $\MP^3 \Nc^2$ is related to the five-dimensional Newton constant as $G = 1/(16 \pi \MP^3 \Nc^2)$. The last term in~\eqref{eq:IHQCDaction} is the Gibbons--Hawking boundary term, which is needed to make the variational problem well defined. However, it does not play a further role in the analysis that follows. 

In addition, $V(\Phi)$ is the dilaton potential, which holographically implements the renormalization group (RG) flow in the model.  
We use the dilaton potential from~\cite{Amorim:2021gat}:
\begin{equation}
\label{eq:IHQCDpotnewmaintext}
\begin{split}
    V(\lambda) &= 12 + V_1 \lambda + V_2 \frac{\lambda^2}{1 + \frac{\alpha_\lambda \lambda}{\lambda_0}} + 3 V_{\text{IR}}e^{- \frac{\lambda_0}{\alpha_\lambda \lambda}} \frac{\lambda^{4/3}}{4 \pi^{8/3}} \sqrt{\log \qty(1 + \frac{\alpha_\lambda \lambda}{\lambda_0})}  \\
    V_1 &= \frac{44}{9\pi^2} \ ,  \qquad V_2 = \frac{4619}{3888\pi^4} \ ,  \qquad \lambda_0 = 8 \pi^2 \ ,
\end{split}
\end{equation}
where $\lambda = e^\Phi$.
This form is designed to reproduce the correct UV and IR behavior of Yang--Mills theory~\cite{Gursoy:2007cb,Gursoy:2007er}. In particular, it matches the pure $\liegroup{SU}{\Nc}{}$ Yang--Mills $\beta$-function up to two-loop order in the UV, while yielding linear confinement in the IR. A detailed discussion of the construction and properties of the model can be found in~\cite{Gursoy:2010fj}.

The fit parameters in~\eqref{eq:IHQCDpotnewmaintext} take the values $\alpha_\lambda = 2.833$ and $V_{\text{IR}} = 7.231$. These were fixed in~\cite{Amorim:2021gat} by fitting the mass spectrum of light mesons, including radially excited states with masses up to approximately $2\,\mathrm{GeV}$, in an extended version of the model that also incorporates quark degrees of freedom. The coefficients $V_1$ and $V_2$ are fixed by the UV matching to the Yang--Mills $\beta$-function.

Apart from the parameters of the potential and the five-dimensional Planck scale $\MP$, the model contains one additional dynamical parameter, namely the energy scale $\Lambda$ that sets the overall scale of the background solutions.

%%%%%%%%%%%%%%%%%%%%%%%%%%%%%%%%%%%%%%%
\subsection{Bulk gravity background}
%%%%%%%%%%%%%%%%%%%%%%%%%%%%%%%%%%%%%%%

In order to solve the background metric in this model, we adopt the {\emph{Ansatz}} 
\begin{equation} \label{eq:metric}
    \dd s^2 =  -e^{2A(r)} \eta_\MN \dd x^M \dd x^N =  e^{2A(r)}\left( -\dd t^2 + \dd {\bf{x}}^2+ \dd r^2 \right) \ ,    
\end{equation}
where $M,N=0,\ldots,4$, and $\eta_\MN = \text{diag}(1,-1,-1,-1,-1)$. Because scalings of the spacetime coordinates (along with a scaling of $r$) can be absorbed by shifts of $A(r)$, its holographic interpretation is the logarithm of the energy scale~\cite{Peet:1998wn}. The $r$-dependence of this  factor is such that it cuts spacetime off dynamically once one enters the deep IR region of large $r$, hence making this model a certain kind of a soft-wall model. In these coordinates the UV boundary is located at $r=0$ and the geometry ends in an IR singularity at $r = \infty$. 

Background fields are solved from Einstein equations for dilaton gravity. After inserting the {\emph{Ansatz}}~\eqref{eq:metric}, these equations boil down to
\bea
    9A'^2 + 3A'' &= & e^{2A}V(\Phi) \\
    3A'^2 - 3A'' &= & \frac{4}{3}\qty(\Phi')^2 \ ,
\eea
where we suppressed the dependence of $A$ and $\Phi$ on the holographic coordinate $r$, so that a prime denotes a derivative with respect to $r$. We follow similar notation for scalar functions throughout the text, unless writing the coordinate dependence explicitly is required for clarity.
The IR and UV asymptotics of the geometry are presented explicitly in Appendix~\ref{app:IHQCD}, and in particular they give an explicit definition for the energy scale $\Lambda$ of the background. 

\subsection{Fluctuations of bulk gravitational fields}\label{sec:gravflucts}

The low-energy spectrum of the Yang--Mills theory contains scalar (spin zero) and tensor (spin two) glueball states. Through the holographic dictionary, these states can be studied in the dual picture by studying the linearized fluctuations of the metric and the dilaton around the gravity background.
A complete analysis of linear gravity perturbations in Einstein-dilaton gravity can be found in \cite{Kiritsis:2006ua}; here we will only discuss the details relevant for the current article. 

After including generic scalar and tensor fluctuations, the metric can be written as 
\be
\label{eq:metricAnsatz}
\dd s^2 = e^{2A(r)}\left[ - \qty(\eta_{\mu\nu} + 2 h \eta_{\mu\nu} - 2 \partial_\mu \partial_\nu E - 2h_{\mu\nu}^{TT}) \dd x^\mu \dd x^\nu +2 \partial_\mu W \dd r\, \dd x^\mu + (1+2 h_r)\dd r^2 \right] \ ,
\ee
where $\eta_\mn = \text{diag}(1,-1,-1,-1)$. The scalar fluctuation wave functions\footnote{For notational simplicity, we mark the coordinate dependence in the argument of all functions by $x$ even though the wave functions may depend on the full coordinate vector $x^\mu$, not only on its length.
} $h(x,r)$, $E(x,r)$, $W(x,r)$, and $h_r(x,r)$, as well as the tensor fluctuation wave functions  $h_{\mu\nu}^{TT}(x,r)$ are infinitesimal. Here $TT$ stands for transverse and traceless, {\emph{i.e.}}, $\partial^\mu h_{\mu\nu}^{TT} = \partial^\nu h_{\mu\nu}^{TT} = h_{\mu}^{TT\,\mu} = 0$.  The linear perturbation of the dilaton $\Phi$ about the background value is denoted by $\varphi(x,r)$. 

Infinitesimal scalar diffeomorphisms $r \mapsto r +\xi_r$, $x^\mu \mapsto x^\mu +\partial^\mu \xi$ act on the scalar perturbations as
\begin{align}\label{eq:xitr1}
\varphi&\mapsto \varphi + \Phi' \xi_r \ ,&\qquad    h_r &\mapsto h_r + A'\xi_r +\xi_r' \ ,& \qquad h \mapsto h +A'\xi_r \ , \\
\qquad W &\mapsto W +\xi_r +\xi' \ ,&\qquad E &\mapsto E +\xi  \ ,&
\label{eq:xitr2}
\end{align}
where we remind that the primes denote derivatives with respect to the holographic coordinate $r$. Consequently, we can define the following diffeomorphism covariant scalar variables 
\be
\label{eq:diffeomorphismscalars}
 \mathcal{F} = h - \frac{A'}{\Phi'} \varphi\ , \qquad \widetilde{\mathcal{F}} = \frac{h}{A'} - W + E' \ , \qquad \mathcal{G} = h' - A' h_r + \frac{4}{9}\Phi' \varphi\ .
\ee

As it turns out, linearly perturbed Einstein equations imply that $\G = 0$~\cite{Kiritsis:2006ua}. Moreover, $\F $ and $ \Tilde{\F}$ are conjugate fields with relations
\be
\label{eq:conjugateequations}
\widetilde{\mathcal{F}}' +3A' \widetilde{\mathcal{F}} - \frac{4}{9}\qty(\frac{\Phi'}{A'})^2 \mathcal{F} = 0 \ , \qquad \partial_\mu\partial^\mu\widetilde{\mathcal{F}} - \frac{4}{9}\qty(\frac{\Phi'}{A'})^2 \mathcal{F}' = 0 \ .
\ee
The tensor mode satisfies
\be \label{eq:tensorflucteq}
 \left(h_{\mu\nu}^{TT} \right)'' + 3 A' \left(h_{\mu\nu}^{TT}\right)' -  \partial_\rho\partial^\rho h_{\mu\nu}^{TT} = 0 \ .
\ee
As a consequence of the background depending only on the holographic coordinate $r$, the spacetime dependence of the equations becomes diagonal in Fourier space. To make this explicit, we write
\begin{equation}
\begin{split}
   \F(x,r) &= e^{-i k \cdot x } \hat f(k,r) \\
   h_{\mu\nu}^{TT}(x,r) &= e^{-i k \cdot x} \epsilon^{TT}_{\mu\nu}(k)\ \hat h^{TT}(k,r)  \ ,
\end{split}
\end{equation}
where the momentum space transverse-traceless projector satisfies $k^{\mu}\epsilon^{TT}_{\mu\nu} = k^{\nu}\epsilon^{TT}_{\mu\nu} = 0 = \epsilon^{TT\,\mu}_{\mu}$. Fluctuation equations~\eqref{eq:conjugateequations} and~\eqref{eq:tensorflucteq} imply that the radial wave functions satisfy
\begin{equation}
\label{eq:gravityeom}
\begin{split}
    \qty(\pd^2_r + 3A' \pd_r + k^2) \hat h^{TT}(k,r) & = 0  \\
     \qty(\pd^2_r + \qty(3A' + 2 \frac{X'}{X})\pd_r + k ^2) \hat f(k,r) & = 0  \qquad , \quad X \equiv 
     \frac{\Phi'}{3A'}   \ .
\end{split}
\end{equation}
Let $\hat f_s$ and $\hat h_s^{TT}$ denote the bulk-to-boundary propagator solutions. We define the boundary conditions for the propagators such that $\hat{h}_s^{TT}(k, r=0) = \hat{f}_s(k, r = 0) = 1$, and impose regularity in the IR. Note that, as usual, the radial wave functions only depend on the momentum through $k^2 = k_\mu k^\mu$. 

Both of the fluctuation equations~\eqref{eq:gravityeom} may be written in Schr\"odinger form where
\begin{equation}
\label{eq:gravityeomschr}
    - \left(\hat h^{TT} \right)'' + V_T \hat h^{TT} = -k^2\hat h^{TT} \ , \qquad -\hat f'' + V_S \hat f = -k^2\hat f
\end{equation}
with potentials 
\begin{equation}
    V_T = \frac{3}{4}A'^2 + \frac{3}{2}\frac{\qty(e^{A})''}{e^{A}}\ , \qquad V_S = V_T + \frac{X''}{X} + 3A'\frac{X'}{X} \ .
\end{equation}
The scalar potential term makes the glueball mass spectra non-degenerate. The spectrum of low-lying glueballs, which is obtained from the normalizable solutions to these equations, agrees well~\cite{Jarvinen:2022gcc} with lattice data  when using the potential~\eqref{eq:IHQCDpotnewmaintext}, even though the potential was not fitted to Yang--Mills data but only to the experimental spectrum of full QCD.

%%%%%%%%%%%%%%%%%%%%%%%%%%%%%%%%%%%%%%%
\subsection{Modeling the proton} \label{sec:protondefs}
%%%%%%%%%%%%%%%%%%%%%%%%%%%%%%%%%%%%%%%

In order to introduce an external proton in the model, we consider a simple approach where the proton is modeled as a fermionic field in the bulk. One is tempted to introduce the proton through a five-dimensional Dirac field in the gravity theory. However, this is problematic: a single five-dimensional Dirac fermion would, in general, model Weyl fermions at the boundary (see, {\emph{e.g.}},~\cite{Foit:2019nsr}), rather than Dirac fermions. In particular this means that the action for a five-dimensional Dirac field is not parity invariant. A parity-invariant model is obtained by considering a combination of two Dirac fields~\cite{Contino:2004vy,Plantz:2018tqf}. Here, we will anyhow start with the action for a single field and discuss below how the model is made parity invariant.

The fermionic action reads
\be\label{eq:particleaction}
    S_F = \N \int \dd^5 x \sqrt{-\det g}\,\L_F + \N \int \dd^4 x \sqrt{-\det g^{(4)}}  \L_{UV}\ , 
\ee
where $\N$ is a normalization factor, and the Lagrangians are given by\footnote{In principle, the derivatives in~\protect\eqref{eq:protonaction} should be promoted to covariant derivatives. In practice, however, the additional connection dependent terms cancel in our explicit computations, including the fluctuation analysis in Appendix~\ref{app:amplitudescouplingsvertices}. This reflects the fact that the variations considered are generated by diffeomorphisms, for which the relevant combinations naturally assemble into Lie derivatives.}
\bea    
    \label{eq:protonaction}
    \L_F & = &  e^{- \Phi}  \qty(\frac{i}{2} \Bar{\Psi} e^N_{\ B} \gamma^B \pd_N \Psi - \frac{i}{2} (\pd_N \Bar{\Psi}) e^N_{\ B} \gamma^B \Psi- \VN(\Phi)\Bar{\Psi} \Psi) \\
    \L_{\text{UV}} & = & e^{- \Phi} (\overline{\Psi}_L \Psi_R + \overline{\Psi}_R \Psi_L) \ .
\eea
The first Lagrangian describes the dynamics of a Dirac fermion $\Psi$. The normalization factor $\N$ is used as a fitting parameter. We also introduced the vielbeins $e^N_{\ B}$, with which the metric is represented as $g^{\MN} = -e^M_{\ A} e^N_{\ B} \eta^\AB$. Here the matrices having upper indices are the inverse metric and inverse of $\eta$. {\color{black}Since the background metric \eqref{eq:metric} is diagonal, background vielbeins} can be taken to be simply $e^N_{\ B} = e^{-A(r)}\delta^N_B$. The five-dimensional gamma-matrices are defined such that the first four matrices are the same as in the four-dimensional algebra, whereas the fifth matrix is $\gamma^r = -i \gamma^5$, with $\gamma^5 \equiv i\gamma^0\gamma^1\gamma^2\gamma^3$. They satisfy the algebra $\anticommutator{\gamma^A}{\gamma^B} = 2 \eta^\AB$. 

The second Lagrangian describes a four-dimensional boundary term, where $\Psi_L, \Bar{\Psi}_L, \Psi_R$, and $\Bar{\Psi}_R$ are the right- and left-handed parts of the full spinors $\Psi, \Bar{\Psi}$ reduced to the boundary, $r=0$. This boundary term is necessary to make the variational problem well-defined \cite{Contino:2004vy} but plays no further role in this article. 

Let us then analyze the fermion spectrum. The equation of motion for fermions is
\begin{equation} \label{eq:fermioneom}
    \qty[i e^N_{\ B} \gamma^B D_N - \frac{i}{2} \qty(\partial_N \Phi) e^N_{\ B} \gamma^B  -  \VN(\Phi)]\Psi = 0 \ .
\end{equation}
Here the covariant derivative is given by $D_N = \pd_N + \frac{1}{8}\omega_{NBC} [\gamma^B, \gamma^C] $, and $\omega_{NBC}$ is the spin-connection term. Its non-zero components are explicitly
\begin{equation}
    \omega_{\mu r \nu} = -\omega_{\mu \nu r} = A'(r) \eta_\mn \ .
\end{equation}
As above, we write a plane-wave {\emph{Ansatz}} 
\begin{equation}
    \Psi(x,r) = \Tilde{\Psi}(p,r) e^{-ip\cdot x}   \ .
\end{equation}
The Fourier components can be decomposed as
\begin{equation}
\label{eq:diracfermion}
    \Tilde{\Psi}(p,r) = \Tilde{\Psi}_L(p,r) + \Tilde{\Psi}_R(p,r) = e^{\frac{\Phi(r)}{2}-2A(r)}\left[\psi_L(r)u_L(p) + \psi_R(r)u_R(p)\right] \ .    
\end{equation}
Here $\Tilde{\Psi}_{L,R}(p,r) = (1\mp  \gamma^5)\Tilde{\Psi}(p,r)/2$, and  $u_{L,R}(p) = (1\mp \gamma^5)u(p)/2$, with upper signs giving the conventions for left-handed spinors and $u(p)$ being the off-shell Dirac spinor. The radial wave functions $\psi_L(r)$ and $\psi_R(r)$ are scalar functions. We found it convenient to include the background dependent exponential factor in the last expression in~\eqref{eq:diracfermion}. 

By using the fact that 
\begin{equation}
\label{eq:spinorrelations}
    \slashed{p} u_R(p) = m_n \,  u_L(p) \ , \qquad \slashed{p} u_L(p) = m_n\, u_R(p) \ ,    
\end{equation} 
with $m_n=\sqrt{p^2}$ for a bound state on-shell and solutions indexed as $n = 0,1,2,\ldots$, 
separating the equation of motion~\eqref{eq:fermioneom} into left- and right-handed modes yields
\begin{equation} \label{eq:fermioncoupled}
    -m_n\,\psi_L = \left[\frac{\dd}{\dd r}  - e^{A}\VN(\Phi)\right]\psi_R \ , \qquad    -m_n\,\psi_R =  \left[-\frac{\dd}{\dd r}  - e^{A}\VN(\Phi)\right]\psi_L \ .
\end{equation}
These coupled equations can in turn be written as second-order equations where the functions are decoupled:
\begin{equation}
\begin{aligned} 
\label{eq:fermiondecoupled}
        m_n^2 \psi_R &=  
        \qty[-\frac{\dd^2}{\dd r^2} + \VR]\psi_R  \ , & \qquad \VR &= e^{2A}\VN(\Phi)^2 + \frac{\dd}{\dd r}\left(e^{A}\VN(\Phi)\right) \ , &
         \\
        m_n^2 \psi_L &=  
        \qty[-\frac{\dd^2}{\dd r^2} + \VL]\psi_L \ ,  & \qquad \VL &= e^{2A}\VN(\Phi)^2 - \frac{\dd}{\dd r}\left(e^{A}\VN(\Phi)\right) \ .& 
\end{aligned}        
\end{equation}
The bound state wave functions are normalized as 
\begin{equation}
    \int \dd r \ \psi_{L,R}^2(r) = 1 \ .
\end{equation}

Normally, the mass spectrum is obtained from the normalized modes of the fluctuation equation. However, in the case of a bulk fermion, there are complications~\cite{Henningson:1998cd}. In general, there is no obvious way to pick a normalizable mode for fermions: the left- and right-handed wave functions are coupled through~\eqref{eq:fermioncoupled}, which appears to prevent one from simultaneously requiring normalizability in both equations of~\eqref{eq:fermiondecoupled}. 

This complication reflects the observation which we pointed out above: a single bulk Dirac fermion is not a good model for the proton, and only models half of the required degrees of freedom. That is, we need to choose whether the action~\eqref{eq:particleaction} models the left- or right-handed component of the proton. We choose it to be the left-handed component, {\emph{i.e.}}, that the spectrum is obtained by solving the Schr\"odinger problem for $\psi_L$ in~\eqref{eq:fermiondecoupled}. Actually, for the specific potential $\VN(\Phi)$ that we will be using here, this appears to be the only reasonable choice: as shown in Appendix~\ref{app:Aasymptotics}, the right-handed wave function $\psi_R$ is UV-normalizable for all solutions, so requiring $\psi_R$ to be normalizable instead would not lead to a discrete spectrum. 

Another way to see the interplay of the left- and right-handed functions is to study more closely their expansions near the boundary, which are solved explicitly in Appendix~\ref{app:Aasymptotics}. Recall that the standard picture in gauge/gravity duality is that each field has two independent solutions near the boundary, with the coefficient of the leading term being identified as the source ({\emph{e.g.}}, a mass) and the coefficient of the subleading term is identified as the vacuum expectation value (VEV) of the dual operator. The coefficients are not linked by the equations of motion, but they become dependent after one imposes regularity ({\emph{e.g.}}, finiteness of the action) in the IR. However, in the case of Dirac bulk fermion, the UV expansions only contain two independent coefficients even though there are two fields, $\psi_L$ and $\psi_R$. These two coefficients can be taken to be the leading coefficient of $\psi_L$ and the leading coefficient of $\psi_R$. Then, we need to choose (see, {\emph{e.g.}},~\cite{Contino:2004vy}), which of these coefficients is interpreted as the source and which is interpreted as the VEV. The above choice, {\emph{i.e.}}, requiring that the mass spectrum is given by the Schr\"odinger equation for $\psi_L$, is equivalent to taking the leading coefficient of $\psi_L$ as the source and the leading coefficient of $\psi_R$ as the VEV.

We identify the proton as the lowest mode of the set of normalizable solutions, {\emph{i.e.}}, we require the lowest mass (with $n=0$) to be equal to the nucleon mass $\mN$, 
\begin{equation}
    m_0\equiv \mN \ .
\end{equation}
The proton bulk eigenfunction is denoted by $\Psi^{0}_p(x,r)$, where the momentum of the proton state satisfies $p^2 = (m_0)^2$. It is understood that the wave functions $\psi_L$ and $\psi_R$ below are always specifically the eigensolutions of the proton state, and we do not indicate this explicitly for notational simplicity.

There is an undefined potential function in~\eqref{eq:protonaction} which enters into the equations of motion of the proton. Following the philosophy in the IHQCD model~\cite{Gursoy:2007cb,Gursoy:2007er}, we want to produce a mass 
spectrum for which squared masses are linear in the excitation number, $m_n^2 \sim n$, to reproduce standard Regge-like behavior. This is obtained by choosing the potential $\VN$ appropriately.
The requirement translates into choosing $\VN(\Phi)$ in (\ref{eq:protonaction}) such that the left-handed Schr\"odinger potential $\VL$ evaluated on the gravity background grows as $\sim r^2$ in the IR. Given the IR asymptotics of the background in Appendix~\ref{app:IHQCD}, this boils down to the requirement 
\be
 \VN(\Phi) \sim e^{\frac{2}{3}\Phi}\Phi^{\frac{1}{4}}
\ee
as $\Phi \to \infty$. We choose a parametrization which satisfies this requirement and is smooth for all values of $\Phi$,
\begin{equation}
\label{eq:protonpotential}
    \VN(\Phi) = \vN \,e^{\frac{2}{3}\Phi}\qty(1 + \Phi^2)^{\frac{1}{8}} \ ,
\end{equation}
where $\vN$ is a constant. Note that it can be matched with the proton mass, but this also requires fixing the overall energy scale, which amounts to determining the value of the parameter $\Lambda$ defined in Appendix~\ref{app:Aasymptotics}.  
We will do this later when comparing the model to the data for the gravitational form factors.

As we pointed out above, the fermion action~\eqref{eq:protonaction} cannot be made parity-invariant. Parity may be restored by adding another bulk Dirac fermion in the model, which is the chiral partner of the $\Psi$ in~\eqref{eq:protonaction}. The action of the chiral partner is obtained from~\eqref{eq:protonaction} by flipping the sign of $\gamma^5$. This means that the $\gamma$-matrix in the holographic direction, $\gamma^r$, flips its sign, and left- and right-handed components are interchanged, {\emph{e.g.}}, $\psi_L \leftrightarrow \psi_R$. The full action with parity restored is given explicitly in Appendix~\ref{app:vertices}, see Eq.~\eqref{eq:symmetrizedaction}.

In the action for the chiral partner, we take the leading coefficient in the UV expansion of the right-handed field (instead of the left-handed one) $\psi_R$ to be the source, as required by parity covariance. However, note that the equations of motion for the chiral partner are obtained from~\eqref{eq:fermioncoupled} by also switching the labels $L$ and $R$. As the switch happens both in the equations and in their boundary conditions, the mass levels remain unchanged.

%%%%%%%%%%%%%%%%%%%%%%%%%%%%%%%%%%%%%%%
%%%%%%%%%%%%%%%%%%%%%%%%%%%%%%%%%%%%%%%
\section{Gravitational form factors}
\label{sec:formfactors}
%%%%%%%%%%%%%%%%%%%%%%%%%%%%%%%%%%%%%%%
%%%%%%%%%%%%%%%%%%%%%%%%%%%%%%%%%%%%%%%

Having set the holographic model, we turn to identifying and computing the gravitational form factors. This requires us to fluctuate the fermionic action \eqref{eq:protonaction} %with respect to the metric 
to extract the matrix element of the energy-momentum tensor with the holographic dictionary. From this matrix element we identify the gluonic form factors $\Ag(t)$ and $\Dg(t)$ in our model. The remaining parameters are then fit into sets of recent lattice data, and results are compared across various studies. 

\subsection{Holographic analysis}

In order to analyze the gravitational form factors of the proton in field theory via holography, we need to compute the matrix element $ \mel{p_2}{T^\mn}{p_1}$ in~\eqref{eq:meldecomposition} using the holographic dictionary. The dictionary for the energy-momentum tensor is the usual: it is dual to the metric of the five-dimensional gravity theory. However, there are some complications about the proton state. First, as we discussed above, the proton is actually modeled through the sum over two five-dimensional Dirac fermions with opposite chiralities. Second, by using the usual definition of the dictionary in terms of generating functionals, it is straightforward to compute correlators of operators, but not scattering amplitudes with on-shell particles.

The first complication is handled by summing over the contributions from the Dirac fermions, see Appendix~\ref{app:vertices} for details. In the main text, we will often only consider contributions from a single fermion for simplicity. As for the second complication, a well-known formalism exists. One may start by considering a three-point function with two insertions of a baryon operator $\mathcal{O}_\Psi$ dual to $\Psi$ and one insertion of $T_{\mu\nu}$. {\color{black}As per holographic dictionary~\cite{Gubser:1998bc}, this is obtained by taking functional derivatives of the on-shell gravity action with respect to the sources, identified as the coefficient of the non-normalizable right-handed fermion boundary modes and the boundary metric, which will be defined explicitly below.}
One then transforms the three point function to Fourier space, which introduces momenta for all three operators. Then one studies the limit where the baryon momenta become on-shell. By the Lehmann--Symanzik--Zimmermann (LSZ) reduction formula, the three-point function becomes, after removing divergences, proportional to the desired matrix element (in Fourier space). On the gravity side, an analogous reduction takes place~\cite{Hoyos:2019kzt}. Following the Gubser--Klebanov--Polyakov dictionary~\cite{Gubser:1998bc}, the generating functional for correlators is obtained in terms of IR-regular fluctuation wave functions, {\emph{i.e.}}, wave functions for which the fluctuated action is IR-finite, and which reduce to desired boundary values at the UV, which act as sources for the field theory operators. However, in the limit where the momentum of a fluctuation wave function becomes on-shell, the IR-regular wave function becomes ill-defined due to a presence of sourceless regular solutions corresponding to the bound states. In the case of Dirac fermions in the setup discussed above, these solutions are given by the normalizable solutions to the (left-handed) Schr\"odinger equation in~\eqref{eq:fermiondecoupled}. This means that as one approaches the on-shell value of the momentum, the fluctuation wave functions diverge. This divergence matches exactly the divergence of the correlator in field theory. Therefore, after applying the LSZ reduction, one recovers the gravity expression for the amplitude, where the fermion fluctuation wave functions are replaced by the coefficients of their divergent terms, {\emph{i.e.}}, the sourceless normalizable bound state wave functions.

\begin{figure}[!htb]
 \centering
 \includegraphics[width=0.8\textwidth]{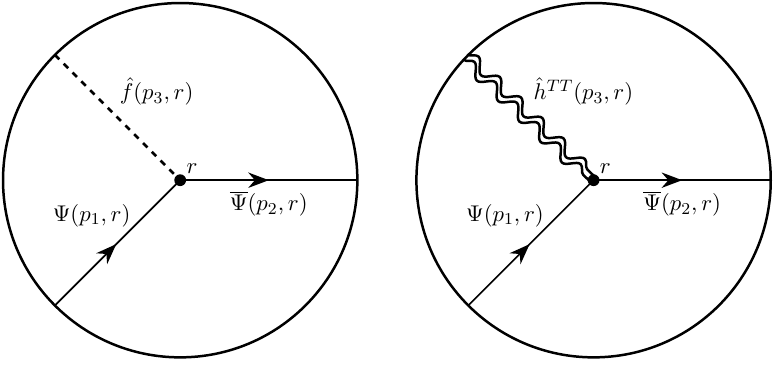}
\caption{Two possible Witten diagrams for three-point functions including Dirac spinors.}
\label{fig:wittendiagrams}
\end{figure}

The variation with respect to the sources in the metric is done as  usual. Varying the gravity action boils down to varying with respect to the {\color{black}boundary} sources for the dynamical fluctuations of the metric $\mathcal{F}$ and $h_{\mu\nu}^{TT}$. To make the definition concrete, we choose a gauge where the fluctuations of the $rr$ and $r\mu$-components of the metric vanish, {\emph{i.e.}}, use the diffeomorphism transformations in~\eqref{eq:xitr1} and~\eqref{eq:xitr2} to set $W=0$ and $h_r=0$. Then the fluctuations of the metric are contained in
\begin{align}
\label{eq:metricgaugefixed}
\dd s^2 &= e^{2A(r)}\left[ - \qty(\eta_{\mu\nu} + h_{\mu\nu}) \dd x^\mu \dd x^\nu + \dd r^2 \right]  & \\
h_\mn &= 2 h \eta_\mn - 2 \partial_\mu \partial_\nu E - 
\partial_{\{\mu} V_{\nu\}}^T - 2h_{\mu\nu}^{TT} \ .&
\label{eq:fulldecomp}
\end{align}
Here we also included the vectorial perturbation $V_\mu^T(x,r)$, where the superscript $T$ indicates that the field is transverse, $\partial^\mu V_\mu^T = 0$. This field is in principle required in order to write a consistent variation formula for the energy-momentum tensor operator.\footnote{By this we mean that there is no regular  solution to the fluctuation equations satisfying the boundary condition~\eqref{eq:grbc} for any $\delta \eta_\mn$, unless we also include the vectorial field.} However, the vectorial field is not a physical propagating mode and its bulk wave functions are trivial ($r$-independent), see, {\emph{e.g.}},~\cite{Kiritsis:2006ua}. It could still contribute to the matrix element $\mel{p_2}{T^\mn}{p_1}$ which we want to compute, but as it turns out, it does not (see Appendix~\ref{app:variation}). Therefore, we will not discuss the contribution from this field further. Given the above gauge choice, we may write the boundary condition involving the variation of the field theory metric $\delta \eta_\mn(x)$, which couples to the energy-momentum tensor operator $T_\mn$, as
\be \label{eq:grbc}
 h_\mn(x,r=0) = \delta \eta_\mn(x) \ .
\ee

In Fourier space, the evaluation of the three-point function therefore boils down to computing the Witten diagrams in Fig.~\ref{fig:wittendiagrams}, with the understanding that the fermion lines represent the on-shell normalizable fluctuation wave functions. The most nontrivial part of the Witten diagram is the vertex between the fermions and the metric fluctuations. This vertex is obtained from the fermion action~\eqref{eq:particleaction}. Note that since our setup does not have fermionic background, the contributions from this action start at second order in the fluctuations. The desired vertex is found from the third-order fluctuation term, which is obtained by fluctuating the metric in the Lagrangian~\eqref{eq:protonaction} (see Appendix~\ref{app:variation} for details). A straightforward computation gives for this term\footnote{We do not attempt to keep track of boundary terms in this computation. As far as we can see, all boundary terms will eventually vanish after we insert the proton wave functions, thanks to their UV asymptotics, which are analyzed in Appendix~\ref{app:IHQCD}.}
\bea
\label{eq:variation}
 S_F^{(3)} & = & \mathcal{N} \int \dd^5x \sqrt{-\det g}\, e^{-\Phi}\bigg[ \frac{\dd \VN(\Phi)}{\dd\Phi} \Bar{\Psi}\Psi \mathcal{F} \frac{\Phi'}{A'} 
 - \frac{2i}{9}
 \frac{\Phi'^2}{A'^2}e^{-A}\left(\Bar{\Psi}'\gamma^r\Psi-\Bar{\Psi}\gamma^r\Psi'\right)\mathcal{F}  \nonumber\\
 & &  + \frac{i}{2}e^{-A}\partial_{\mu}\left(\Bar{\Psi}'\gamma^\mu\Psi-\Bar{\Psi}\gamma^\mu\Psi'\right) \widetilde{\mathcal{F}} + \frac{i}{2}e^{-A}\left( \partial^\mu\Bar{\Psi}\gamma^\nu\Psi-\Bar{\Psi}\gamma^\mu\partial^\nu\Psi\right)h_{\mu\nu}^{TT}  \bigg]   \ , \label{eq:diffeomorphismvariation}
\eea
where prime denotes a derivative with respect to the holographic coordinate $r$ as before, and the gauge-invariant combinations of the fluctuation wave functions of the metric: $\mathcal{F}$, $\widetilde{\mathcal{F}}$, and $h_{\mu\nu}^{TT}$ are defined in Section~\ref{sec:gravflucts}. Because the additional gauge-invariant wave function $\mathcal{G}$ vanishes after imposing the Einstein equations, we have omitted the terms involving this field. {\color{black}We remark that the diffeomorphism-covariant coupling of the fermion fields to the scalar wave functions $\mathcal{F}$ and $\widetilde{\mathcal{F}}$ in~\eqref{eq:variation}, obtained here by varying a consistent gravity model implementing a dynamical soft-wall, differs drastically from the vertices used previously in simple soft-wall models~\cite{Mamo:2019mka}.}

Taking into account the above considerations, we can finally write down the correspondence between the field theory amplitude and gravity action. We first define the on-shell vertex as
\be \label{eq:S3on-shell}
 S^{(3)}_{F\,\mathrm{o-s}}(p_2,p_1)[\delta \eta_{\mu\nu}]  = S^{(3)}_F\Big|_{\bar\Psi = \bar \Psi^0_{p_2}\,,\ \Psi = \Psi^0_{p_1}\,,\ h_{\mu\nu}|_\mathrm{bdry} = \delta \eta_{\mu\nu}  } \ ,
\ee
where $S^{(3)}_F$ is given in (\ref{eq:diffeomorphismvariation}), and the proton wave functions\footnote{To be precise, we need to insert the wave functions for both Dirac fermions that model the proton in a way which conserves parity, see Appendix~\ref{app:vertices}.} are as defined in Section~\ref{sec:protondefs}. Note that this expression is a function of the proton momenta and a functional of the boundary variation $\delta \eta_\mn$. The source of the dilaton variation $\varphi$ is set to zero. The above discussion leads to the following identification
\be
 \int \dd^4x\, e^{ip\cdot x }  \mel{p_2}{T^\mn(x)}{p_1} = (2\pi)^4\delta^{(4)}(p-p_1+p_2)\mel{p_2}{T^\mn(0)}{p_1} \propto \int \dd^4x\, e^{ip\cdot x } \frac{\delta S^{(3)}_{F\,\mathrm{o-s}}}{\delta \left(\delta\eta_\mn(x)\right)} \ .
\ee
Therefore taking the inverse Fourier transform gives simply
\bea
    \mel{p_2}{T^\mn(x)}{p_1}  & \propto &  \frac{\delta S^{(3)}_{F\,\mathrm{o-s}}}{\delta \left(\delta\eta_\mn(x)\right)}  \ .
    \label{eq:correlators}
\eea
The proportionality coefficient, which arises from the holographic LSZ reduction, is not needed for our purposes: it turns out that it can be absorbed into other parameters of the setup.

In order to evaluate the functional derivative in~\eqref{eq:correlators}, it is useful to compute the on-shell vertex~\eqref{eq:S3on-shell} explicitly. We sketch here the main points, see Appendix~\ref{app:vertices} for details. Using~\eqref{eq:fulldecomp} and~\eqref{eq:grbc}, we can solve the boundary values of the metric fluctuations
\begin{equation}
\label{eq:hhmunutt}
    \begin{split}
        h(x,r=0) & =  \frac{1}{6}(\eta^\mn - \pd^\mu \pd^\nu/\pd^2)  
        \delta \eta_\mn(x)\\
        E(x,r=0) & =   \frac{1}{6}\frac{1}{\pd^2}\qty(\eta^\mn - 4 \pd^\mu \pd^\nu/\pd^2)\delta \eta_\mn(x)
        \\
        h_{\mn}^{TT}(x,r=0) & =   \left[-\frac{1}{2}\qty(\delta^\alpha_\mu - \pd^\alpha \pd_\mu/\pd^2)\qty(\delta^\beta_\nu - \pd^\beta \pd_\nu /\pd^2) \right.\\
        &\qquad \left. + \frac{1}{6}\qty(\eta^\ab - \pd^\alpha \pd^\beta/\pd^2)\qty(\eta_\mn - \pd_\mu \pd_\nu/\pd^2)\right] \delta\eta_\ab(x) \\
        &\equiv - \frac{1}{2}\varepsilon_\mn^{TT,\ab}\delta \eta_\ab(x) \ ,
          \end{split}
\end{equation}
where the operator inverse $1/\pd^2$ is defined as
\be
 \frac{1}{\pd^2}f(x) \equiv \int \frac{\dd^4p}{(2\pi)^4}e^{-ip\cdot x} \frac{1}{-p^2} \hat f(p)  \ ,
\ee
for any function $f$ and its Fourier transform $\hat f$. Because we only turn on sources for the fluctuations of the metric and not for the dilaton, from~\eqref{eq:diffeomorphismscalars} we see that the boundary values of the diffeomorphism invariant fluctuation $\mathcal{F}$ and the scalar metric fluctuation $h$ match,
\be
 h(x,r=0) =  \mathcal{F}(x,r=0) \ .
\ee
Moreover, because the term involving $\widetilde{\mathcal{F}}$ in~\eqref{eq:variation} vanishes after we have inserted the solutions for the wavefunctions of the proton state (see Appendix~\ref{app:vertices}), we only need the solutions for the fluctuations $\mathcal{F}$ and $h_\mn^{TT}$. They may be written as
\begin{align}
\label{eq:Fhmunu}
 \F(x,r) &= \int \frac{\dd^4p}{(2\pi)^4} \hat f_s(p,r)e^{-ip\cdot x} \int \dd^4 y\, e^{ip\cdot y } h(y,r=0) & \nonumber\\
&= \int \frac{\dd^4p}{(2\pi)^4} \hat f_s(p,r)e^{-ip\cdot x} \frac{1}{6}(\eta^\mn - p^\mu p^\nu/p^2)\int \dd^4 y\, e^{ip\cdot y } \delta \eta_{\mn}(y) \\
h_\mn^{TT}(x,r) &= \int \frac{\dd^4p}{(2\pi)^4} \hat h_s^{TT}(p,r)e^{-ip\cdot x} \int \dd^4 y\, e^{ip\cdot y } h_\mn^{TT}(y,r=0) & \nonumber\\
& = \int \frac{\dd^4p}{(2\pi)^4} \hat h_s^{TT}(p,r)e^{-ip\cdot x} \left(- \frac{1}{2}\hat\epsilon_\mn^{TT,\ab}(p)\right) \int \dd^4 y\, e^{ip\cdot y } \delta \eta_{\ab}(y) \ ,
\label{eq:hmunuTT}
\end{align}
where $\hat f_s$ and $\hat h_s^{TT}$ are the bulk-to-boundary propagators defined in Sec.~\ref{sec:gravflucts} and
\begin{equation}
   \hat \epsilon_\mn^{TT,\ab}(p) =  \qty(\delta^\alpha_\mu - p^\alpha p_\mu/p^2)\qty(\delta^\beta_\nu - p^\beta p_\nu /p^2) - \frac{1}{3}\qty(\eta^\ab - p^\alpha p^\beta/p^2)\qty(\eta_\mn - p_\mu p_\nu/p^2) \ .
\end{equation}
Inserting these and the proton wavefunctions in~\eqref{eq:variation} yields (see Appendix~\ref{app:vertices})
\begin{equation}
\begin{split}
     S^{(3)}_{F\,\mathrm{o-s}}(p_2,p_1)[\delta \eta_{\mu\nu}]= &\ \N \int \dd r  \Bigg[ 2 \frac{\Phi'}{A'} e^A \frac{\dd \VN(\Phi)}{\dd \Phi} \ub(p_2) \psi_L \psi_R u(p_1) \hat f_s(k,r) \frac{1}{6}(\eta^\mn - k^\mu k^\nu/k^2)  \\
    &-\frac{4}{9} \qty(\frac{\Phi'}{A'})^2 \ub(p_2) (\psi_L' \psi_R - \psi_L \psi_R') u(p_1) \hat f_s(k,r) \frac{1}{6}(\eta^\mn - k^\mu k^\nu/k^2) \\
    &+\left. (\psi_R^2 + \psi_L^2) \ub(p_2) \frac{P^{\{ \alpha}\gamma^{\beta \}}}{2} u(p_1) \hat{h}_s^{TT}(k,r) \left(- \frac{1}{2}\hat\epsilon_\ab^{TT,\mn}(k)\right) \right] \\
   & \times \int \dd^4 y\, e^{i k\cdot y } \delta \eta_{\mn}(y)  \ ,
\end{split}    
\end{equation}
where $k=p_2-p_1$.

Taking the functional derivative in~\eqref{eq:correlators} therefore gives
\begin{equation}
    \begin{split}
    &\mel{p_2}{T^\mn(x)}{p_1}  \\
        & = \frac{2}{6} \widetilde{\N} \int \dd r \frac{\Phi'}{A'} e^A \frac{\dd \VN(\Phi)}{\dd \Phi} \psi_R \psi_L \hat f_s(k,r) \frac{1}{k^2} \ub(p_2)\qty(k^2 \eta^\mn - k^\mu k^\nu) u(p_1)e^{ik\cdot x} \\
       &\quad - \frac{2}{27} \widetilde{\N} \int \dd r \qty(\frac{\Phi'}{A'})^2 (\psi_R \psi_L' - \psi_R' \psi_L) \hat f_s(k,r) \frac{1}{k^2} \ub(p_2)\qty(k^2 \eta^\mn - k^\mu k^\nu) u(p_1) e^{ik\cdot x} \\
        &\quad - \frac{1}{2} \widetilde{\N} \int \dd r\,  (\psi_R^2 + \psi_L^2) \hat h_s^{TT}(k,r) \ub(p_2) \frac{P^{ \{ \mu} \gamma^{\nu \}}}{2}u(p_1)e^{ik\cdot x}  \\
       &\quad +\frac{1}{6}\widetilde{\N} \int \dd r\,   (\psi_R^2 + \psi_L^2) \hat h_s^{TT}(k,r) \ub(p_2)  \mN \frac{1}{k^2}\qty(k^2 \eta^\mn - k^\mu k^\nu)u(p_1)e^{ik\cdot x}  \ ,
    \end{split}
\end{equation}
where $k=p_2-p_1$ and the dimensionless coefficient $\widetilde{\N}$ is the product of $\N$ and the dimensionful proportionality constant in the relation~\eqref{eq:correlators}.  Comparing with the decomposition of the matrix element,
\be
\mel{p_2}{T^\mn(x)}{p_1} = \Bar{u}(p_2) \qty(\Ag(t)\frac{P^{\{ \mu}\gamma^{\nu\}}}{2}+ \Bg(t) \frac{i P^{\{\mu}\sigma^{\nu \}\alpha} k_\alpha}{4\mN} + \Dg(t) \frac{k^\mu k^\nu - \eta^\mn k^2}{4 \mN})u(p_1)e^{ik\cdot x} \ ,
\ee
where the Mandelstam variable is defined as $t = k^2$, we may directly identify the form factors, which are interpreted to be the gluonic contributions, because our gravity background is dual to purely gluonic Yang-Mills theory. Noting that there is no structure corresponding to $\Bg$ it is identically zero, while the other two form factors read
\bea
\label{eq:finalformfactors}
        \Ag(t) & = & - \frac{\widetilde{\N}}{2} \int \dd r\, \hat h_s^{TT}\qty(\sqrt{-t},r) (\psi_R^2 + \psi_L^2) \\
        \Dg(t) & = &-\frac{8 \mN \widetilde{\N}}{6 t}\Bigg[\frac{1}{2} \mN \int \dd r\, \hat h_s^{TT}\qty(\sqrt{-t},r) (\psi_R^2 + \psi_L^2) + \int \dd r\, \hat f_s\qty(\sqrt{-t},r)   \nonumber\\
        & & \times \qty(\qty(\frac{\Phi'}{A'}) e^A \frac{\dd \VN(\Phi)}{\dd\Phi} \psi_R \psi_L - \frac{2}{9}\qty(\frac{\Phi'}{A'})^2\qty(\psi_R \psi_L' - \psi_R' \psi_L))\Bigg] \ .
        \label{eq:finalDfactor}
\eea

%%%%%%%%%%%%%%%%%%%%%%%%%%%%%%%%%%%%
\subsection{Fitting the model parameters to data}
\label{subsec:fitting}
%%%%%%%%%%%%%%%%%%%%%%%%%%%%%%%%%%%%

We first need to fix the remaining model parameters. There are three free parameters left, which are the overall normalization $\widetilde{\N}$, the coefficient $\vN$ multiplying the proton potential in \eqref{eq:protonpotential}, and the characteristic energy scale $\Lambda$. 

In order to understand the fitting procedure, it is useful to analyze the units of energy in our setup. There are actually two natural ways to assign the units. They can be defined through the following transformations, which leave the action invariant (possibly up to an overall factor). The first one,
\begin{align} \label{eq:transell}
 x^\mu &\mapsto   x^\mu/\Lambda_\ell \ , \qquad r \mapsto  r/\Lambda_\ell \ , \qquad V(\Phi) \mapsto V(\Phi) \Lambda_\ell^2 \ , \qquad M_\mathrm{P} \mapsto M_\mathrm{P} \Lambda_\ell \ , \\
 \qquad \VN(\Phi) &\mapsto \VN(\Phi) \Lambda_\ell \ , 
\end{align}
is perhaps the canonical choice. It
leaves the components of the metric unchanged but alters the potentials $V$ and $\VN$. Therefore it also sets the units of the AdS radius, which is determined from the UV normalization of $V$ (see Appendix~\ref{app:Aasymptotics}), to the inverse of energy as expected. We could also assign units to the fermionic field $\Psi$, but these are not needed for our purposes, and can be absorbed in the dimension of $\widetilde{\N}$. The second one is obtained by transforming the components of the metric instead, $g_\mn \mapsto g_\mn \Lambda_s^2$, or in terms of the scale factor $A$,
\begin{align} \label{eq:transs}
 x^\mu &\mapsto   x^\mu/\Lambda_s \ , \qquad r \mapsto  r/\Lambda_s \ , \qquad A \mapsto A +\log \Lambda_s 
\end{align}
with all other fields and parameters unchanged. That is, the potentials and the AdS radius are also unchanged and therefore counted as dimensionless.  Note that the four-momenta transform as $k^\mu \mapsto k^\mu \Lambda_s$, and masses of bound states also transform linearly in $\Lambda_s$. 

Since the line element is invariant under~\eqref{eq:transs}, it is immediate that the action is invariant. In particular, the transformation $e^N_{\ B} \mapsto e^N_{\ B}/\Lambda_s$ of the vielbein cancels the transformation of the derivative in the fermion action~\eqref{eq:protonaction}. This latter transformation may appear unnatural, since it does not respect the expected dimensions of the potentials and the AdS radius. However, the fact that the potentials are unchanged under this transformation also means that it is a symmetry of the solutions of the equation of motion. The parameter in the background solution which corresponds to this symmetry is the scale factor $\Lambda$. Therefore, when analyzing the solutions numerically, we can change the value of $\Lambda$ through the symmetry~\eqref{eq:transs}. The freedom corresponding to the first transformation~\eqref{eq:transell} is fixed by choosing the UV normalization of the potential $V(\Phi)$ such that the AdS radius $\ell$ equals one.

The basic idea of the fit is that the coefficient of the fermion potential, $\vN$, is determined by the proton mass, whereas $\widetilde{\N}$ and $\Lambda$ are determined by fitting data for the form factor $\Ag(t)$. Note that thanks to the symmetry~\eqref{eq:transs}, changing $\Lambda$ corresponds to scaling the $t$-dependence of the form factors, while changing $\widetilde{\N}$ changes the normalization, which makes fitting the form factor rather straightforward. However, note that the fits to the proton mass and the form factor are not independent: Since the proton mass changes under the symmetry~\eqref{eq:transs} while the potential $\VN(\Phi)$ does not, the fit is affected by the value of $\Lambda$. Conversely, the choice of the coefficient $\vN$ affects the shape of the holographic result for $\Ag(t)$ and therefore modifies the fit of the form factor. Therefore, we carry out the two fits iteratively until convergence is found.

We fit our $\Ag(t)$ to the lattice data for the gluonic contributions to $\A(t)$ obtained in \cite{Shanahan:2018pib,Hackett:2023rif}. In the lattice study of \cite{Shanahan:2018pib} the pion mass is rather high, $m_\pi = 450$ MeV, while in \cite{Hackett:2023rif} the mass is closer to the physical value, $m_\pi = 170$ MeV. The gluon and quark contributions to the total form factors are not independently conserved, and hence, are renormalization scale-dependent~\cite{Polyakov:2018zvc}. Both sets of lattice data used in the fit have been computed using the modified minimal subtraction scheme at $\mu = 2$ GeV. Only data for $\Ag(t)$ are used in the fit, since these datasets are significantly more uniform than the currently available results for $\Dg(t)$. An additional motivation for this choice is that fixing $\Ag(t)$ allows remaining proton properties, which are determined through $\Dg(t)$, to be predicted within the model. Had $\Dg(t)$ been fitted directly instead, the resulting mechanical properties would largely be imposed by the fit itself rather than emerging as genuine model predictions. At each step of the iterative fitting procedure we do a least-squares fit of $\Ag(t)$ to the data~\cite{Shanahan:2018pib,Hackett:2023rif}, taking into account the error estimates in the data.

We pick an initial value $\vN \approx 1$ and fit $\Lambda$ repeatedly to lattice data using the proton mass and other parameters. This procedure converges quickly after only a few iterations and yields the parameters $\vN \approx 0.217,$ $ \widetilde{\N} \approx -0.523$, and $\Lambda \approx 147$~MeV.
This sets $\Ag(0) \approx 0.523$, and as required, $\mN \approx  0.938$~GeV.

We then compare our fit to earlier studies.
In many studies ({\emph{e.g.}}, \cite{Mamo:2019mka,Hackett:2023rif, Shanahan:2018pib,Burkert:2018bqq}) 
one encounters so-called multipole fits for the form factor profiles. They are common mainly due to their reasonably good reproduction of results and analytic simplicity. They are of the form
\begin{equation}
\label{eq:multipole}
    \Apol(t) = \frac{\A_0}{\qty(1-\frac{t}{M_\A^2})^{\alpha_\A}} \ ,
\end{equation}
where $M_\A$ is an energy scale and $\alpha_\A$ is an exponent which becomes relevant at high $t$. Similar profiles are encountered also for $\Dpol(t)$ with analogous parameters $\D_0, M_\D$, and $\alpha_\D$. Common choices for the parameters $\alpha_i$, with $i=\A$ or $i=\D$, are $\alpha_i = 2$ (dipole fit) and $\alpha_i = 3$ (tripole fit).

We also plot a parametrized form motivated by the soft-wall–based study of~\cite{Mamo:2019mka}. Choosing $\A_0=0.58$, $M_{\A} = 1.124\,\mathrm{GeV}$, and $\alpha_{\A} = 2$ yields a curve that matches the corresponding result shown there to good precision. For the lattice data of Hackett et al.~\cite{Hackett:2023rif}, the form factor $\Ag(t)$ has been fit with parameters $\A_0 = 0.501 \pm 0.027, \ M_{\A} = 1.262 \pm 0.018 $ GeV, and $\alpha_{\A} = 2$. The $\Ag(t)$ of Shanahan et al. from \cite{Shanahan:2018pib} has parameters $\A_0 = 0.58, \ M_{\A} = 1.13$ GeV, and $\alpha_{\A} = 2$.

\begin{figure}[!htb]
    \centering
    \includegraphics[width=0.85\textwidth]{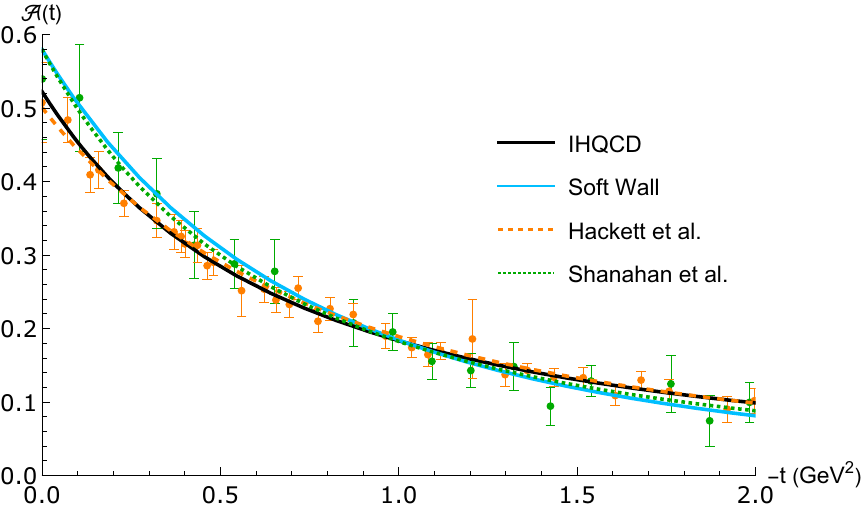}
    \caption{Results for the form factor $\Ag(t)$. Green data points and the green curve (Shanahan et al.) are based on lattice data from \cite{Shanahan:2018pib}. Orange data points and the orange curve (Hackett et al.) are based on lattice data from \cite{Hackett:2023rif}. The soft-wall-based result is from~\cite{Mamo:2019mka}. The black solid curve is our fit.} 
    \label{fig:akfinal}
\end{figure}

We compare our lattice data-fit holographic profile of $\Ag(t)$ (black solid curve) to the multipole fits listed above in Fig. \ref{fig:akfinal}. The lattice data marked in green is from \cite{Shanahan:2018pib}, and data marked in orange from \cite{Hackett:2023rif}.
Note that our IHQCD result is computed directly from the model rather than a dipole formula.
The different fits of $\Ag(t)$ are comparable for a range of $t$-values, and our holographic numerical result overlaps with the lattice fit of Hackett et al.~\cite{Hackett:2023rif}.

%%%%%%%%%%%%%%%%%%%%%%%%%%%%%%%%%%%%
\subsection{Prediction for $\Dg(t)$}
%%%%%%%%%%%%%%%%%%%%%%%%%%%%%%%%%%%%

\begin{figure}[!htb]
    \centering
    \includegraphics[width=0.85\textwidth]{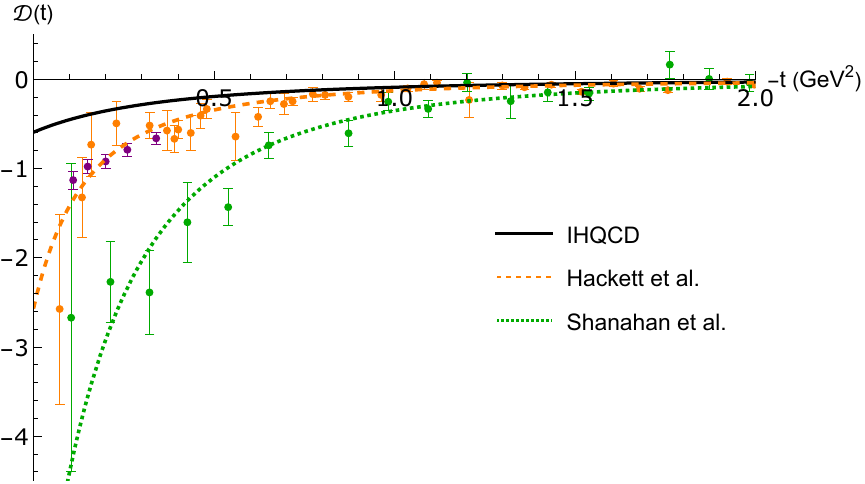}
    \caption{Our prediction (solid black curve) compared to lattice results for the form factor $\Dg(t)$. Green data points and the green curve (Shanahan et al.) are based on lattice data from \cite{Shanahan:2018nnv} and the orange ones (Hackett et al.) are based on lattice data from \cite{Hackett:2023rif}. 
    Dashed curves are multipole fits (\ref{eq:multipole}) to the data points using the values given below the respective equation.}
    \label{fig:dkfinal}
\end{figure}

As one readily checks from (\ref{eq:finalDfactor}), there are no parameters left to fit for $\Dg(t)$. We thus compare the results of the holographic form factor $\Dg(t)$, predicted by the lattice data fit of $\Ag(t)$, to lattice results (see Fig.~\ref{fig:dkfinal}).

For the lattice data of Hackett et al.~\cite{Hackett:2023rif}, the form factor $\Dpol(t)$ has been fit with parameters $\D_0 = -2.57 \pm 0.84, \ M_\D = 0.538 \pm 0.065$ GeV, and $\alpha_D = 2$. The $\D(t)$ result by Shanahan et al.~\cite{Shanahan:2018nnv} has parameters $\D_0 = -7.08, M_\D = 0.763$ GeV, and $\alpha_\D = 3$. These results are compared to our holographic results (solid black curve).

There is considerable variation between different $\Dg(t)$-profiles in Fig.~\ref{fig:dkfinal}. However, the error bars are also large, and the IHQCD results and the lattice results of~\cite{Hackett:2023rif}, with nearly physical pion mass, are relatively close in the range where data is available. Note that, assuming that the difference in the lattice results between the two references mainly arises from the pion mass, the values of $\Dg(t)$ increase with decreasing pion mass. Therefore, extrapolation to the physical pion mass could bring the lattice result even closer to our holographic result.

Of special interest in many studies has been the D-term, {\emph{i.e.}}, the value $\D(0)$ of the form factor at zero momentum transfer \cite{Polyakov:2018zvc}. Looking at \eqref{eq:finalDfactor}, there could in principle be a divergence in $\D(t)$ as $t \to 0$, unless an exact cancelation takes place between the two integrals. We find this cancelation numerically and show it to take place analytically in Appendix~\ref{subapp:cancelationofthedivergence}. Therefore, the model predicts a finite $\D_g(0)$ with the value $\D_g(0) \approx -0.595$. As a reminder, this is only the gluonic contribution to the full $\D(0)$. As an indirect comparison, a top-down holographic study of \cite{Fujita:2022jus, Sugimoto:2025btn} using the Witten--Sakai--Sugimoto found the total $\D(0)\approx -2.05$.

%%%%%%%%%%%%%%%%%%%%%%%%%%%%%%%%%%%%%%%
%%%%%%%%%%%%%%%%%%%%%%%%%%%%%%%%%%%%%%%
\section{Mechanical structure}
\label{sec:mechanicalstructure}
%%%%%%%%%%%%%%%%%%%%%%%%%%%%%%%%%%%%%%%
%%%%%%%%%%%%%%%%%%%%%%%%%%%%%%%%%%%%%%%

In this section we present holographic results for the mechanical properties of the proton.
By definition, gravitational form factors encode information about matrix elements of the energy-momentum tensor of the proton. In particular, the form factor $\D(t)$ is related to spatial deformations of the proton and provides access to its mechanical properties. These include quantities such as pressure and shear force distributions, as well as proton radii defined from moments of the corresponding distributions.

Such mechanical properties have been studied using lattice QCD~\cite{Shanahan:2018pib,Hackett:2023rif}, phenomenological extractions based on DVCS data~\cite{Burkert:2018bqq,Burkert:2021ith}, and holographic approaches~\cite{Mamo:2019mka}. In the case of experimental data, these quantities are obtained indirectly and necessarily rely on modeling assumptions and theoretical constraints, since pressure and shear distributions are not directly observable. It has been emphasized that present data alone are insufficient to determine mechanical properties of the proton in a model-independent manner, and that phenomenological extractions necessarily involve assumptions about functional forms and extrapolations~\cite{Kumericki:2009uq,Kumericki:2019ddg}.

In the following section we always first present the derivation of the mechanical property of interest in a canonical way with total form factors $\A$, $\B$, and $\D$, and specialize to using only the gluonic contribution when computing this quantity, either by using our model or by using lattice data. The results extracted from DVCS data \cite{Burkert:2021ith} are for the total form factor $\D(t)$ including quark contributions.

%%%%%%%%%%%%%%%%%%%%%%%%%%%%%%%%%%%%%%%
\subsection{Pressure and shear force distributions of a proton}
\label{sec:fundamentalquantities}
%%%%%%%%%%%%%%%%%%%%%%%%%%%%%%%%%%%%%%%

The spatial components of the energy-momentum tensor $T_{ij}$ can be decomposed using the pressure $p(\rho)$ and shear $s(\rho)$ profiles of the proton into a trace and traceless part \cite{Polyakov:2018zvc} 
\begin{equation}
\label{eq:spatialenergymomentum}
    T_{ij}(\boldsymbol{\rho}) = p(\rho) \delta_{ij} + s(\rho) \qty(\frac{\rho_i\rho_j}{\rho^2} - \frac{1}{3} \delta_{ij}) \ ,
\end{equation}
where $\boldsymbol{\rho}$ is the position vector from the center of the proton, $\rho_i$ are its components, and $\rho$ denotes its length. These pressure and shear force distributions may be computed directly from the $\D$ form factor. We obtain spatial information by transforming into coordinate space, 
\begin{equation}
\label{eq:Dfouriertransform}
    \widetilde{\D}(\rho) = \int \frac{\dd^3 K}{2 \mN (2\pi)^3} e^{-i \boldsymbol{K} \cdot \boldsymbol{\rho}}\D(-\boldsymbol{K}^2) \ .
\end{equation}
The pressure and shear profiles can be expressed in terms of $\widetilde{\D}(\rho)$ as~\cite{Polyakov:2018zvc}
\begin{equation}
    \label{eq:shearandpressure}
    p(\rho) = \frac{1}{3} \frac{1}{\rho^2} \dv[]{}{\rho} \qty( \rho^2 \dv[]{}{\rho} \widetilde{\D}(\rho))  \ , \qquad s(\rho) = - \frac{1}{2} \rho \dv[]{}{\rho} \qty( \frac{1}{\rho} \dv[]{}{\rho} \widetilde{\D}(\rho)) \ .
\end{equation}
Note that carrying out the transform in~\eqref{eq:Dfouriertransform} requires the knowledge of the form factor at arbitrarily large $|t|$. However, it is challenging to compute $\Dg(t)$ numerically at large enough $|t|$ in our model. Therefore, we match the numerical result along its first two derivatives with an asymptotic expansion to $\Dg(t)$ at approximately $-t = 10 $ GeV$^2$, and use the expansion to extrapolate to high $|t|$. We use the form $\Dg(t) = -\frac{D_1}{(t/\text{GeV}^2)^3} + \frac{D_2}{(t/\text{GeV}^2)^5} - \frac{D_3}{(t/\text{GeV}^2)^7}$ in the extrapolation, which agrees the asymptotic behavior predicted by QCD counting rules \cite{Tanaka:2018wea,Brodsky:1973kr}. The values of parameters obtained from matching are $D_1 \approx 2.17$, $D_2 \approx 124$, and $D_3 \approx 3610$.

\begin{figure}[!htb]
    \centering
    \includegraphics[width=\textwidth]{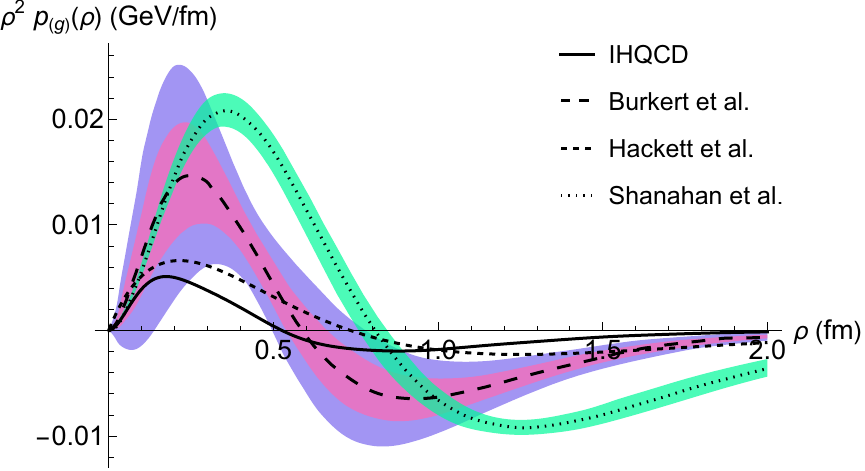}
    \caption{
    Contributions to the pressure distributions~\eqref{eq:shearandpressure} of a proton, as given by different models. The long-dashed black curve, as well as the violet and pink bands arise from a dipole fit analysis of DVCS data from Burkert et al.,~\cite{Burkert:2018bqq}, including contributions both from quarks and from gluons. The violet band shows the error estimate of~\cite{Burkert:2018bqq}, compared to a future projection (pink band). The solid curve is our result for the gluonic contribution. The dotted curve with green error band is a lattice fit result  for the gluonic contribution taken from \cite{Shanahan:2018nnv}. The result labeled Hackett et al. is computed by us from the form of $\Dg(t)$ fitted to lattice data in \cite{Hackett:2023rif}.}
    \label{fig:pressureinproton}
\end{figure}

\begin{figure}[!htb]
    \centering
    \includegraphics[width=\textwidth]{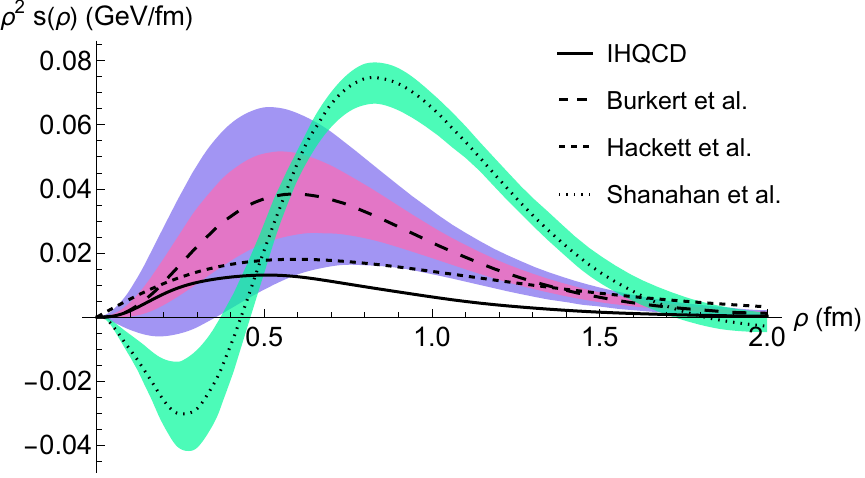}
    \caption{Contributions to the shear force distributions~\eqref{eq:shearandpressure} of a proton, as given by different models. Notation as in Fig.~\protect\ref{fig:pressureinproton}.}
    \label{fig:shearinproton}
\end{figure}

In Figs.~\ref{fig:pressureinproton} and~\ref{fig:shearinproton} we compare the holographic result for the gluonic contribution to the pressure and shear force distributions to two different lattice results (obtained from the tripole fit of \cite{Shanahan:2018nnv}, and the dipole fit of \cite{Hackett:2023rif}) and the result obtained from dipole analysis of DVCS data~\cite{Burkert:2021ith}, which also includes the quarks. The holographic results are similar to the DVCS fit and to the recent lattice result of \cite{Hackett:2023rif},
while differing significantly from the lattice result of~\cite{Shanahan:2018nnv} with higher pion mass. However, the DVCS result for both distributions is somewhat higher than the results from IHQCD and from \cite{Hackett:2023rif}, which could reflect the additional quark contribution included in the DVCS result. Note that the normalizations differ by roughly a factor of two, which is similar to the difference of the gluonic and total form factors $\Ag(t)$ and $\A(t)$: as seen from Fig.~\ref{fig:akfinal}, $\Ag(0)\approx 0.5$ in these models whereas the total form factor satisfies $\A(0)=1$.

For the shear force, our holographic results are again similar to the dipole fit result of~\cite{Burkert:2018bqq} and the result of lattice analysis in \cite{Hackett:2023rif}, but significantly different from lattice result of \cite{Shanahan:2018pib}. The lattice result from the fit of \cite{Hackett:2023rif} raises quicker than our prediction at low $\rho$, but its maximum is located at approximately the same radius as the maxima of the holographic result and the DVCS fit of Burkert et al.~\cite{Burkert:2018bqq}. The behavior at low $\rho$ is 
connected to the power with which $-\D(t)$ decays at high momentum transfer. The decay behavior based on the multipole fit of \cite{Burkert:2018bqq} is approximately $\D(t) \sim 1/(-t)^3$, which again conforms with predictions of QCD counting rules \cite{Brodsky:1973kr,Tanaka:2018wea}.

One may define tangential and normal forces acting inside a proton from the pressure and shear profiles \cite{Polyakov:2018zvc}. They are defined with respect to a principal axis $\rho$, $\theta$, or $\phi$, and the sign of their value represents squeezing or stretching in this direction. By a classical analogy, multiplying pressure by the area element allows us to compute forces acting inside a proton. The profiles can be computed from the decomposition of the spatial part of the energy momentum tensor \eqref{eq:spatialenergymomentum}. For example, projecting $T^\ij$ onto the radial unit vector $\hat{\rho}$ yields 
\begin{equation}
   p_\rho(\rho)\equiv  \hat{\rho}^i T_\ij \hat{\rho}^j = \frac{2}{3} s(\rho) + p(\rho) \ .
\end{equation}
We similarly obtain the angular contribution, 
\begin{equation}
\label{eq:lowerdimensionalforces}
   p_\theta(\rho)= p_\phi(\rho)= - \frac{1}{3}s(\rho) + p(\rho) \ .
\end{equation}
By angular isotropicity, $p_\theta = p_\phi$. 
By a classical analogy such pressures may be related to forces inside a proton. Hence, the magnitude of the normalized profiles describes the relative magnitudes of forces acting inside a proton.

\begin{figure}[!htb]
    \centering
    \includegraphics[width=\textwidth]{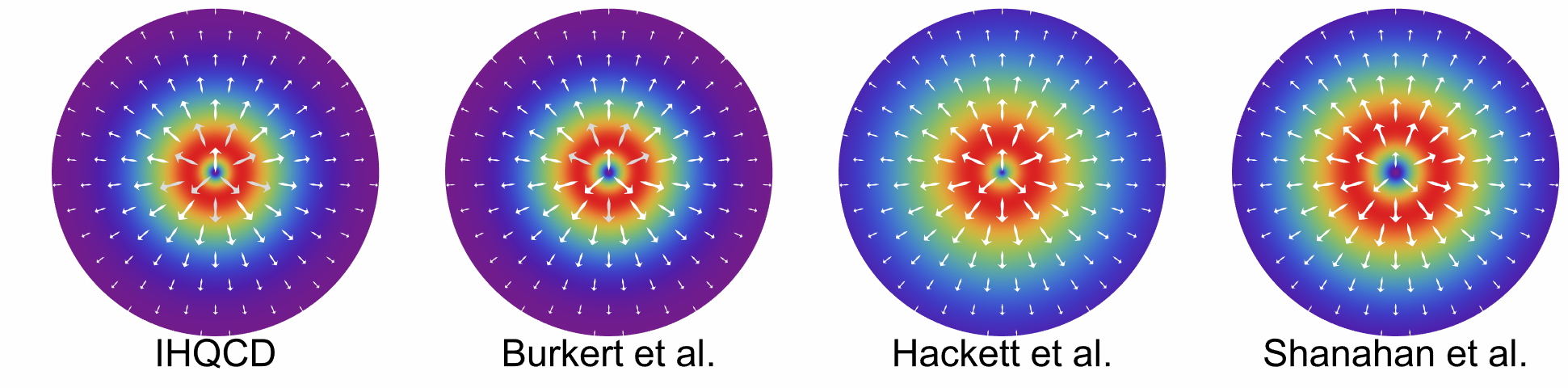}
    
    \vspace{5mm}
    
    \includegraphics[width=\textwidth]{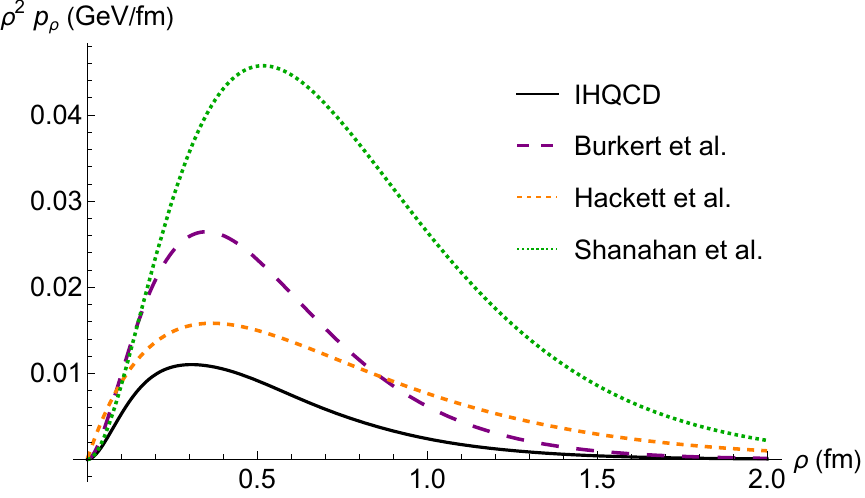}
    \caption{In the upper top panel we show the normalized  
    normal pressures $\rho^2 p_{\rho}/|\rho^2 p_{\rho}|_\text{max}$ for each model. In the bottom panel we show the corresponding absolute
    normal pressures $\rho^2 p_{\rho} \ (\rho\leq 2\,\rm{fm})$ within a proton. We computed the DVCS and lattice results using the fits given in~\cite{Burkert:2018bqq,Hackett:2023rif,Shanahan:2018pib}, see text for details.}
    \label{fig:radialpressures}
\end{figure}

\begin{figure}[!htb]
    \centering
    \includegraphics[width=\textwidth]{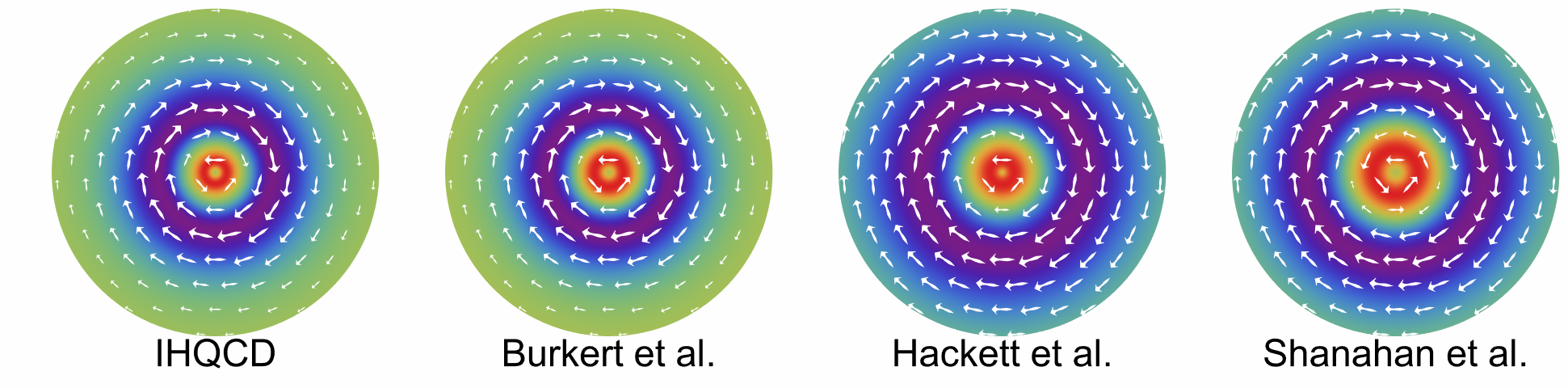}
    
    \vspace{5mm}
    
    \includegraphics[width=\textwidth]{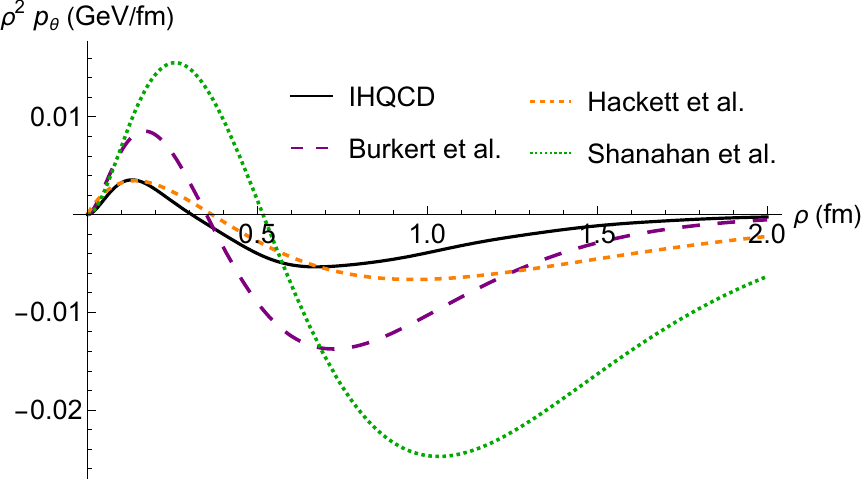}
    \caption{In the top panel we show the normalized  
    tangential pressures $\rho^2 p_{\theta}/|\rho^2 p_{\theta}|_\text{max}$ in different models. In the bottom panel we show the corresponding absolute  
    tangential pressures $\rho^2 p_{\theta} \ (\rho\leq 2\,\rm{fm})$ within a proton. We computed the DVCS and lattice results using the fits given in~\cite{Burkert:2018bqq,Hackett:2023rif,Shanahan:2018pib}, see text for details.}   \label{fig:angularpressures}
\end{figure}

We present visualizations of the normal and tangential pressure distributions for the four models in Figs.~\ref{fig:radialpressures}~and~\ref{fig:angularpressures}.  We computed the profiles labeled Burkert et al. using the $\D(t)$ of the phenomenological fit in \cite{Burkert:2021ith}. We also computed the lattice curves (Hackett et al. and Shanahan et al.) from similar fits to the gluonic contribution $\Dg(t)$ given in \cite{Hackett:2023rif} and \cite{Shanahan:2018nnv}, respectively. The IHQCD result was obtained directly from the results for the pressure and the shear force given above.

The shapes of the distributions between different models are rather similar both in Fig.~\ref{fig:radialpressures} and in Fig.~\ref{fig:angularpressures}. However there are also some differences; For example, in IHQCD and DVCS profiles the angular pressures are more concentrated near the core of the proton, while for the lattice profiles they extend further from the core. Similar to Figs.~\ref{fig:pressureinproton} and~\ref{fig:shearinproton}, the normalization of the DVCS result (Burkert et al.) is significantly higher than in IHQCD and the lattice result with nearly physical pion mass (Hackett et al.), which may reflect the additional quark degrees of freedom present in the DVCS result. 
Radial pressures are positive in all models, being especially large right outside the core, decreasing towards larger radius. As shown in Fig.~\ref{fig:angularpressures}, the angular pressure changes sign at around $\rho \simeq 0.3\text{--}0.5$ fm.

%%%%%%%%%%%%%%%%%%%%%%%%%%%%%%%%%%%%%%%
\subsection{Proton radius}\label{sec:energydensitiesandradii}
%%%%%%%%%%%%%%%%%%%%%%%%%%%%%%%%%%%%%%%

For a given distribution $X$, we may define the proton to have a radius $\rho_X = \sqrt{\lara{\rho^2}_X}$. For example, the (expectation value of the square of the) mass-radius of the proton is defined with respect to the energy density, which is given in momentum space as 
\begin{equation}
     \tilde\epsilon(t) \equiv T^{00}(t) = \mN \qty[\A(t) - \frac{t}{4\mN^2} \qty(- \B(t) + \D(t))]\ , 
\end{equation}
where $\mN$ is the mass of the nucleon. To compute the radius, we Fourier transform back to coordinate-space. With spherical symmetry and $t = - K^2$,
\begin{equation}
\label{eq:energydensityFT}
    \epsilon(\rho) = \frac{\mN}{2\pi^2} \int_0^\infty \dd K \frac{\sin (K\rho) K}{\rho} \qty[\A(-K^2) + \frac{K^2}{4\mN^2} \qty(- \B(-K^2) + \D(-K^2))] \ .
\end{equation}
Then the expectation value of the squared radius is
\begin{equation}
    \lara{\rho^2}_{\text{mass}} = \frac{\int \dd^3 \rho \ \rho^2 \epsilon(\rho)}{\int \dd^3 \rho \  \epsilon(\rho)} \ .
\end{equation}

We collect results for the radius $\rho_{\text{mass}}=\sqrt{\lara{\rho^2}_{\text{mass}}}$ for IHQCD, dispersive analysis~\cite{Cao:2024zlf,Cao:2025dkv}, and lattice fits in Table~\ref{tab:radii}.\footnote{Note that Burkert et al.~\cite{Burkert:2018bqq} only fits the form factor $\D(t)$, which is not enough to compute $\rho_{\text{mass}}$.}
The value of $\rho_\text{mass}$ based on \cite{Shanahan:2018nnv} was computed by us using the tripole fit from this reference and with the assumption that $\Bg \approx 0$. The values of~\cite{Cao:2024zlf,Hackett:2023rif} as well as the error estimates were taken directly from the references.

Another formula for a radius can be derived using the radial pressure $p_\rho$. With this radial distribution of forces, one may define the mean squared mechanical radius \cite{Polyakov:2018zvc},
\begin{equation}
    \lara{\rho^2}_{\text{mech}} = \frac{\int \dd^3 \rho \  \rho^2 p_\rho(\rho)}{\int \dd^3 \rho \ p_{\rho}(\rho)} \ .
\end{equation}
We collect results for $\rho_{\text{mech}}$ as well from IHQCD, dispersive analysis, lattice, and DVCS fits in Table~\ref{tab:radii}. The value of $\rho_\text{mech}$ for Burkert et al.~\cite{Burkert:2021ith} was computed by us from the DVCS fit given in the reference. The other values and their error estimates are taken directly from the references.
\begin{table}[!htb]
    \centering
    \begin{tabular}{|c|c|c|}
    \hline
        Model & $\rho_{\text{mech}}$ (fm)  & $\rho_{\text{mass}} $ (fm) \\
         \hline \hline
        IHQCD   &   0.626   &  0.648  \\
        \hline
        Hackett et al.~\cite{Hackett:2023rif}   &  $0.900 \pm 0.110$   &  $0.820 \pm 0.07$   \\
        \hline
        Shanahan et al.~\cite{Shanahan:2018nnv}   &  $0.755 \pm 0.007$   &  1.08   \\
        \hline 
        Duran et al.~\cite{Duran:2022xag}   &  -    & $0.472 \pm 0.042$      \\
        \hline \hline
        Burkert et al.~\cite{Burkert:2021ith}   &  $0.635^{\,*}$    & -      \\
        \hline
        Cao et al.~\cite{Cao:2024zlf}  &  $0.72^{+0.09\,*}_{-0.08}$    & $0.70^{+0.03\,*}_{-0.04}$      \\
        \hline
        
    \end{tabular}
    \caption{{\color{black}Proton mechanical and mass radii from different analyses. IHQCD denotes our results. Burkert et al. is based on a fit to DVCS data, Duran et al. on a generalized parton distribution and vector meson dominance analysis of near threshold $J/\psi$ photoproduction data, Cao et al. on a dispersive analysis, while Shanahan et al. and Hackett et al. are lattice results. The results marked with an asterisk also contain quark contributions. See the text for details.}}
    \label{tab:radii}
\end{table}

\begin{figure}[!htb]
    \centering
    \includegraphics[width=0.8\textwidth]{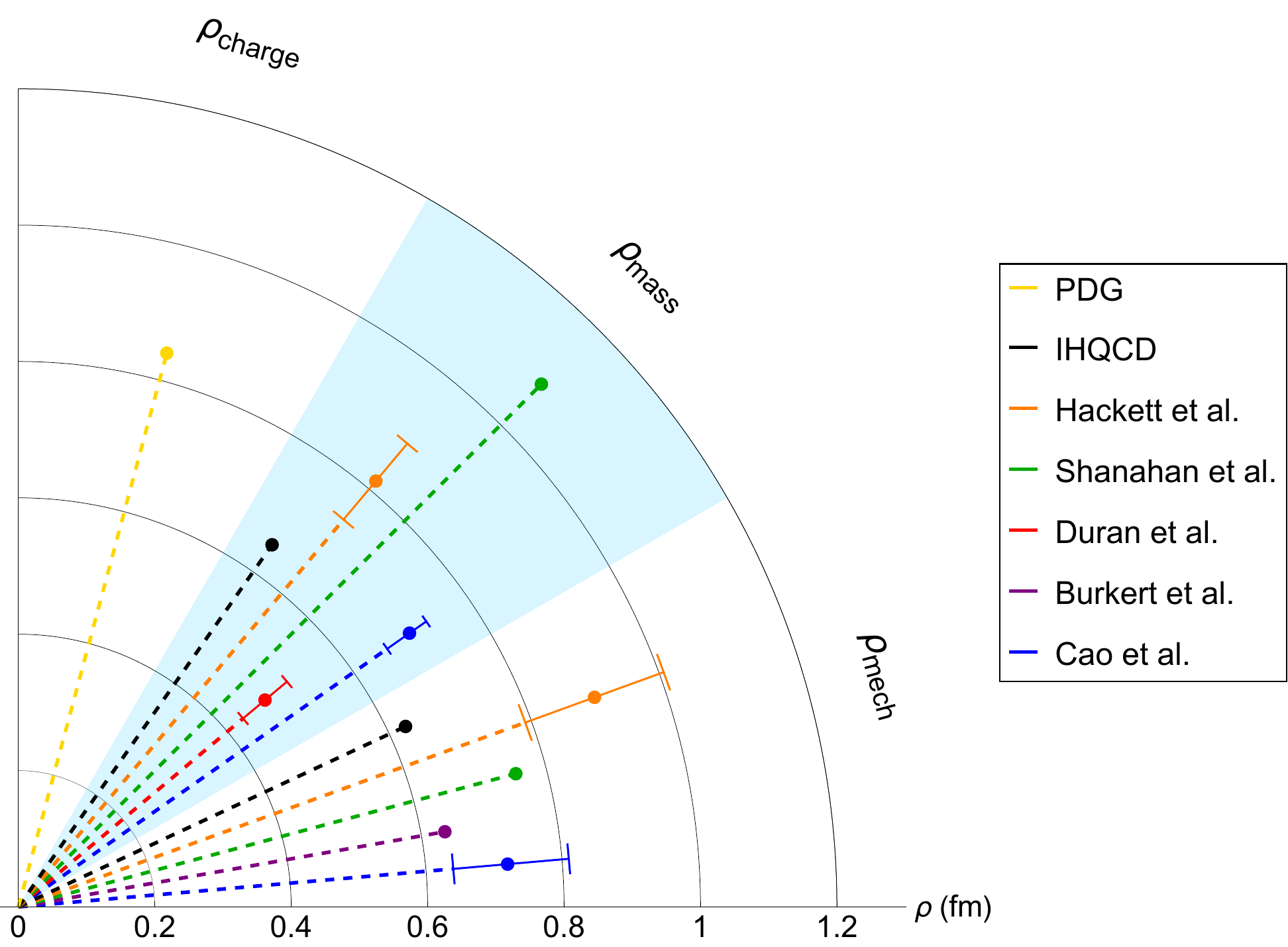}
    \caption{{\color{black}Various determinations of the proton radius. For reference, the value $\rho_\text{charge}=0.841\,\mathrm{fm}$ extracted from muonic hydrogen spectroscopy~\cite{ParticleDataGroup:2024cfk,Antognini:2013txn} is also indicated.}}
    \label{fig:radiusplot}
\end{figure}

For a more visual approach, we plot these radii, computed from the fitted profiles of $\D(t)$ and $\Dg(t)$'s, in Fig. \ref{fig:radiusplot}. In the sector of the mechanical and mass radii, the results are shown in the same order as in Table~\ref{tab:radii}. The result for charge radius is the value listed in PDG \cite{ParticleDataGroup:2024cfk}
and the work \cite{Antognini:2013txn}. The different radii $\rho_{\text{mech}}$, $\rho_{\text{mass}}$, and $ \rho_{\text{charge}}$ are comparable across models considered in this work. For a recent overview of proton radii in different models, see~\cite{Goharipour:2025yxm}.

%%%%%%%%%%%%%%%%%%%%%%%%%%%%%%%%%%%%%%%
%%%%%%%%%%%%%%%%%%%%%%%%%%%%%%%%%%%%%%%
\section{Conclusions and outlook}\label{sec:conclusions}
%%%%%%%%%%%%%%%%%%%%%%%%%%%%%%%%%%%%%%%
%%%%%%%%%%%%%%%%%%%%%%%%%%%%%%%%%%%%%%%

We analyzed the structure of the proton using holographic methods, focusing on the gluonic contribution to its gravitational form factors. Since improved holographic QCD is specifically constructed to capture the dynamics of pure Yang–Mills theory, it provided a natural background geometry for this study. In this work, the proton was modeled as a bulk Dirac fermion propagating in a fixed background geometry, which should be viewed as an effective description rather than a fully realistic proton model.

Unlike many earlier holographic analyses, we formulated a consistent gravitational framework in which the relevant form factors were computed directly from the holographic dictionary. The calculation combined scalar and tensor gravitational fluctuations, fermionic bound states, and the diffeomorphism-induced couplings required by general covariance. The remaining free parameters were fixed by performing a least-squares fit of the form factor $\Ag(t)$ to lattice data~\cite{Shanahan:2018pib,Hackett:2023rif}, which in turn yielded a nontrivial prediction for $\Dg(t)$.  In the limit $t \to 0^-$, the structures contributing to a possible divergence $\Dg(t)$ were shown to cancel exactly; hence, we obtained a finite prediction for the gluonic contribution of $\D(0)$ to be $\D_g(0)\approx -0.595$.

Using the extracted form factors, we studied mechanical properties of the proton by exploiting the relation between the form factors and spatial deformations. This allowed us to define pressure and shear force distributions, which were analyzed and compared with existing lattice and phenomenological studies~\cite{Shanahan:2018pib,Hackett:2023rif,Burkert:2018bqq,Burkert:2021ith,Mamo:2019mka}. We emphasize that such mechanical profiles are not directly measured experimentally, but are inferred under additional assumptions and constraints, and should therefore be interpreted with appropriate care~\cite{Kumericki:2009uq,Kumericki:2019ddg}. The resulting proton radius estimates in our model were found to be $\rho_{\text{mech}} \approx 0.63$~fm and $\rho_{\text{mass}} \approx 0.65$~fm.

Our present holographic setup describes the proton using bulk Dirac fermions propagating in a gravity background dual to pure Yang--Mills theory. While the fermionic fields carry spin, the minimal gravitational couplings included here do not generate the tensor structure required for a nonvanishing Pauli-type GFF, and consequently $\Bg(t)=0$ in this model. The vanishing of $\Bg(t)$ therefore reflects the simplicity of the effective action \eqref{eq:protonaction} rather than the absence of spin degrees of freedom.

Nevertheless, based on the lattice results of \cite{Hackett:2023rif}, a nonzero $\Bg(t)$ would contribute a small correction to energy density at low momentum transfer. Incorporating additional couplings that generate a nonvanishing $\Bg(t)$ would be required for a fully realistic description and is left for future work. 

An important direction for future work would be the inclusion of quark degrees of freedom~\cite{Jarvinen:2011qe,Deng:2025fpq}. This would allow access to quark contributions to the gravitational form factors and their associated mechanical properties, as well as to their interplay with the gluonic sector. Lattice results for quark gravitational form factors are already available~\cite{Hackett:2023rif}. Incorporating flavor dependence is essential for addressing spin-dependent structure, but achieving this in a controlled holographic setting remains technically challenging and is the subject of ongoing work~\cite{Alvares:2011wb,Fang:2018vkp,Cai:2022omk,Zhang:2022uin,Jarvinen:2025mgj,Shen:2025yrn}. In this context, forthcoming measurements at the Electron-Ion Collider are expected to provide valuable constraints on spin-dependent observables~\cite{Becattini:2024uha,Becattini:2025twu}.

Another future direction would be to develop the model from using the effective five-dimensional fermionic action towards more realistic modeling of the proton. It is known that the proper way of describing baryons in dual gravity model at large $N_c$ is via soliton configurations of gauge fields. Such solitons have indeed been constructed in the extension of the IHQCD model to include flavors, the V-QCD model~\cite{Jarvinen:2022mys,Jarvinen:2022gcc}. Analyzing the form factors by using these solutions would therefore be an important extension of this article.

More broadly, gravitational form factors are closely connected to other nonperturbative observables describing hadron structure, including generalized parton distributions and generalized distribution amplitudes, and indirectly to structure functions. Clarifying these connections within holographic frameworks, and confronting them with lattice and experimental constraints, remains an important open problem for future investigation.

\section*{Acknowledgements}

We thank Feng-Kun Guo, Carlos Hoyos, Risto Paatelainen, and Roman Zwicky for discussions. N.~J. was supported in part by the Research Council of Finland through grant no. 354533 and the Centre of Excellence in Neutron-Star Physics (project 374062).

\appendix

%%%%%%%%%%%%%%%%%%%%%%%%%%%%%%%%%%%%%%%
%%%%%%%%%%%%%%%%%%%%%%%%%%%%%%%%%%%%%%%
\section{Details on the computation of the form factors}\label{app:amplitudescouplingsvertices}
%%%%%%%%%%%%%%%%%%%%%%%%%%%%%%%%%%%%%%%
%%%%%%%%%%%%%%%%%%%%%%%%%%%%%%%%%%%%%%%

In this appendix we compute and discuss in detail the fluctuation of the fermion action and the resulting vertices. In particular, we find the relevant terms from which to identify the gravitational form factors and discuss how to make the setup parity invariant. 

%%%%%%%%%%%%%%%%%%%%%%%%%%%%%%%%%%%%%%%
\subsection{Varying the fermion action}
\label{app:variation}
%%%%%%%%%%%%%%%%%%%%%%%%%%%%%%%%%%%%%%%
We start from the fermion action 
\be
\label{eq:SFdef}
 S_F = \mathcal{N} \int \dd^5x \sqrt{-\det g}  e^{- \Phi}  \qty(\frac{i}{2}\Bar{\zeta} e^N_{\ B} \gamma^B \partial_N \zeta-\frac{i}{2} \left(\partial_N\Bar{\zeta}\right) e^N_{\ B} \gamma^B \zeta - \VN(\Phi)\Bar{\zeta} \zeta),
\ee
which agrees with the proton action in~\eqref{eq:particleaction} and~\eqref{eq:protonaction} but (for reasons that will become clear below) we now denote the fermion field by $\zeta$.

There is a complication regarding the fermion field $\zeta$. Since we do not introduce a background for the fermions it is a probe, and hence interpreted as a fluctuation. So, the action~\eqref{eq:SFdef} is already second order in fluctuations. One might expect that with this setup, it is not necessary to consider transformations of $\zeta$ under diffeomorphisms, because the transformation rules are 
\be
 \zeta \mapsto \zeta + \zeta'\xi_r + \partial_\mu \zeta \partial^\mu \xi \ , \qquad   \Bar{\zeta} \mapsto \Bar{\zeta} + \Bar{\zeta}'\xi_r + \partial_\mu \Bar{\zeta} \partial^\mu \xi \ ,
\ee
where the variation terms are of second order in fluctuations and therefore suppressed with respect to the original fields. However, one needs to be careful, because these rules imply that the second-order terms in fluctuations could change the third-order fluctuation term, which we want to compute. To avoid this change from happening, {\emph{i.e.}}, in order to remove the mixing of terms of different order under diffeomorphisms, we introduce a fermion field which is invariant up to second order in fluctuations under diffeomorphisms, 
\be  \label{eq:Psilarel}
 \Psi = \zeta -\frac{\zeta'}{A'}h- \partial_\mu \zeta \partial^\mu E \ , \qquad \zeta = \Psi
+\frac{\Psi'}{A'}h + \partial_\mu \Psi \partial^\mu E \ .
\ee
Note that this definition is not unique: one could replace the term proportional to $h$ by an analogous term involving $h_r$, for example. However, this ambiguity does not affect the final result, because the subleading fluctuation terms in~\eqref{eq:Psilarel} can be interpreted as a variation of the fermion field, and their effect on the third-order fluctuation term vanishes after one imposes the fermion equation of motion~\eqref{eq:fermioneom}. This also means that strictly speaking, for our purposes, introducing a gauge invariant fermion field would not be necessary, but we nevertheless find it convenient to do so.

In order to compute the third-order fluctuation term, we write first the second-order term from~\eqref{eq:SFdef} in terms of the covariant field $\Psi$. After this, we insert the fluctuations both for the fermions from~\eqref{eq:Psilarel} and for the metric from~\eqref{eq:metricAnsatz} in the action and compute the third-order term in fluctuations. The choice of vielbeins $e^M_{\ B}$ is not unique. A natural choice is, including only  scalar fluctuations up to linear terms, 
\begin{align}
 e^r_{\ r} &= e^{-A}\left(1- h_r\right)\ , \qquad  &  e^\mu_{\ r} &= \frac{1}{2}(\pd^\mu W) e^{-A} \ ,& \nonumber \\ \label{eq:vielbeins}
 e^r_{\ \mu} &= - \frac{1}{2}(\pd_\mu W) e^{-A} \ , \qquad & e^{\mu}_{\ \nu} &= e^{-A}\delta^\mu_\nu\left(1- h \right)+ e^{-A}\partial^\mu\partial_\nu E \ .
\end{align}
The expressions for $e^r_{\ r}$ and $e^\mu_{\ \nu}$ are the only natural covariant choices here, but the expressions for the other components, $e^\mu_{\ r}$ and $e^r_{\ \mu}$, could also be chosen differently. However, as it turns out, all choices give the same result for the fermion-scalar vertex that we want to compute.
A direct yet quite tedious computation gives 
\bea
\label{eq:S3F}
 S_F^{(3)} & = & \mathcal{N} \int \dd^5x \sqrt{-\det g}\, e^{-\Phi}\bigg[\frac{\Phi'}{A'}\frac{\dd \VN(\Phi)}{\dd\Phi} \Bar{\Psi}\Psi \mathcal{F}- \frac{2i}{9}
 \frac{\Phi'^2}{A'^2}e^{-A}\left(\Bar{\Psi}'\gamma^r\Psi-\Bar{\Psi}\gamma^r\Psi'\right)\mathcal{F}  \nonumber\\
 & & \quad + \frac{i}{2}e^{-A}\partial_{\mu}\left(\Bar{\Psi}'\gamma^\mu\Psi-\Bar{\Psi}\gamma^\mu\Psi'\right) \widetilde{\mathcal{F}} + \frac{i}{2}e^{-A}\left( \partial^\mu\Bar{\Psi}\gamma^\nu\Psi-\Bar{\Psi}\gamma^\mu\partial^\nu\Psi\right)h_{\mu\nu}^{TT}  \bigg] \ ,
\eea
where we used Einstein equations both for the background as well as for the fluctuations, and neglected boundary terms. 

Einstein equations also imply that $\widetilde{\mathcal{F}}$ is the conjugate field of $\mathcal{F}$, as shown in (\ref{eq:conjugateequations}). Moreover, they set the third field to zero, $\mathcal{G} = 0$. A second-order action for the dynamical fields, {\emph{i.e.}}, $\mathcal{F}$ and $h_{\mu\nu}^{TT}$, can be found by expanding the gravity action (see, {\emph{e.g.}},~\cite{Kiritsis:2006ua})
\be
 S_\mathrm{IHQCD}^{(2)} = -\MP^3\Nc^2 \int \dd^5 x \sqrt{-\det g}\left[ \frac{\Phi'^2}{A'^2}g^{MN} \partial_M \mathcal{F} \partial_N \mathcal{F} + g^{MN} \partial_M h_{\mu\nu}^{TT} \partial_N h^{TT\,\mu\nu}\right]\ .
\ee

Finally, let us check the coupling of the helicity-one fields in gravity to the fermion. We do not expect that these non-propagating fields contribute to the gravitational form factors, but this coupling term provides an additional consistency check of our computations.
We follow the conventions of~\cite{Kiritsis:2006ua}, and include the fields $A_\mu^T$ and $V_\mu^T$ in the fluctuated metric,
\begin{align}
\label{eq:metricvector}
\dd s^2 = e^{2A(r)}\Big[ &- \qty(\eta_{\mu\nu} + 2 h \eta_{\mu\nu} - 2 \partial_\mu \partial_\nu E - \partial_{\{\mu}V_{\nu\}}^T - 2h_{\mu\nu}^{TT}) \dd x^\mu \dd x^\nu \nonumber\\
&+2 \left( \partial_\mu W +A_\mu^T\right)\dd r \dd x^\mu + (1+2 h_r)\dd r^2 \Big] \ .
\end{align}
Here $\partial^\mu V_\mu^T= 0 = \partial^\mu A_\mu^T$.
In principle, we also need to add the vectorial contribution in~\eqref{eq:Psilarel} to make sure that $\Psi$ remains invariant also in spin-one variations of the coordinates, by writing
\be
\zeta = \Psi
+\frac{\Psi'}{A'}h + \partial_\mu \Psi \partial^\mu E - \partial^\mu\Psi V_\mu^T \ .
\ee
However, the contributions from this extra term in the fluctuation of the fermion action again trivially vanish. 
Therefore, it is enough to consider the variations of the vielbeins in~\eqref{eq:SFdef}, extending the expressions~\eqref{eq:vielbeins} to include the helicity-one fields. Here, it is important to use the ``symmetric'' definition of the components $e^\mu_{\ r}$ and $e^r_{\ \mu}$ where both of them are nonzero, because otherwise the covariant derivatives of the fermions, which we have omitted in the action~\eqref{eq:SFdef}, would contribute to the vertex. 
Dropping boundary terms, the coupling between the fermions and the helicity-one fields becomes 
\be
\label{eq:S3FV}
 S_{FVA}^{(3)}  =  \mathcal{N} \int \dd^5x \sqrt{-\det g}\, e^{-\Phi}\frac{i}{4}\left(- \Bar{\Psi}\gamma^\mu\Psi'+\Bar{\Psi}'\gamma^\mu\Psi+\Bar{\Psi}\gamma^r\partial^\mu\Psi-\partial^\mu\Bar{\Psi}\gamma^r\Psi\right)\left(A_\mu^T-{V_\mu^T}'\right) \ .
\ee
Using the equations of motion for the helicity-one fields  from~\cite{Kiritsis:2006ua}, this term vanishes,\footnote{The vanishing of the term requires assuming that the gravity fluctuations are massive ({\emph{i.e.}}, the four-dimensional Fourier modes of the wave functions have a nonzero mass), but as we shall see, this is the case for the wave functions contributing to the form factors. 
} as expected.

%%%%%%%%%%%%%%%%%%%%%%%%%%%%%%%%%%%%%%%
\subsection{Restoring parity}
\label{app:vertices}
%%%%%%%%%%%%%%%%%%%%%%%%%%%%%%%%%%%%%%%

As we pointed out in Section~\ref{sec:protondefs}, restoring parity invariance of the action requires adding two Dirac fermions in the bulk having opposite parities. 
Doing this explicitly, the total fermion action reads
\begin{equation}\label{eq:symmetrizedaction}
    S = \N \int \dd^5 x \sqrt{-\det g}\,\L_F + \N \int \dd^4 x \sqrt{-\det g^{(4)}}  \L_{UV} \ ,
\end{equation}
with
\bea
    \L_F & = &  e^{- \Phi}  \left[\frac{i}{2} \qty(\Bar{\Psi}_1 e^\mu_{\ B} \gamma^B \pd_\mu \Psi_1 + \Bar{\Psi}_1 e^r_{\ B} \gamma^B \pd_r \Psi_1) \right. \\
    &  & \left. +\frac{i}{2} \qty((\pd_\mu \Bar{\Psi}_1) e^\mu_{\ B} \gamma^B \Psi_1 + (\pd_r \Bar{\Psi}_1) e^r_{\ B} \gamma^B \Psi_1)- \VN(\Phi)\Bar{\Psi}_1 \Psi_1  \right] \nonumber \\
    &  & + e^{- \Phi}  \left[\frac{i}{2} \qty(\Bar{\Psi}_2 e^\mu_{\ B} \gamma^B \pd_\mu \Psi_2 - \Bar{\Psi}_2 e^r_{\ B} \gamma^B \pd_r \Psi_2) \right. \\
    &  & \left. - \frac{i}{2} \qty((\pd_\mu \Bar{\Psi}_2) e^\mu_{\ B} \gamma^B \Psi_2 - (\pd_r \Bar{\Psi}_2) e^r_{\ B} \gamma^B \Psi_2)- \VN(\Phi)\Bar{\Psi}_2 \Psi_2\right] \nonumber\\
    \L_{\text{UV}} & = &  e^{- \Phi} (\overline{\Psi}_{1,L} \Psi_{1,R} + \overline{\Psi}_{1,R} \Psi_{1,L}) +e^{- \Phi} (\overline{\Psi}_{2,R} \Psi_{2,L} + \overline{\Psi}_{2,L} \Psi_{2,R}) \ ,
\eea
where $\Psi_1$ and $\Psi_2$ are the chiral partner Dirac fields. Note that there are sign changes in the $r$-derivative terms, effectively obtained by changing the sign of $\gamma^r$ between the two fields, as we explained in the main text.

Imposing that the fermion fields represent the proton state with momentum $p^\mu$, the expression in~\eqref{eq:diracfermion} generalizes to
\begin{align}
\Psi_1(x,r) &= e^{\frac{\Phi(r)}{2}-2A(r)}\left[\psi_L(r)u_L(p) + \psi_R(r)u_R(p)\right]e^{-ip \cdot x } & \\
\Psi_2(x,r) &= e^{\frac{\Phi(r)}{2}-2A(r)}\left[\psi_R(r)u_L(p) + \psi_L(r)u_R(p)\right]e^{-ip \cdot x } \ , & 
\end{align}
with the understanding that $\psi_L$ and $\psi_R$ in $\Psi_1$ are the proton wave functions specified in the main text, so that $\psi_L$ is UV-normalizable. That is, the change of the handedness between $\Psi_1$ and $\Psi_2$ amounts to just changing the left- and right-handed solutions,
\begin{equation}
    \psi_{R/L} \ \leftrightarrow \ \psi_{L/R} \ .
\end{equation}

Following these definitions, let us simplify all terms of $S^{(3)}_F$ in~\eqref{eq:S3F}. The total fluctuation for a two-fermion action reads
\bea
\label{eq:S3Fbothfermions}
 S_F^{(3)} & = & \mathcal{N} \int \dd^5x \sqrt{-\det g}\, e^{-\Phi}\bigg[\frac{\Phi'}{A'}\frac{\dd \VN(\Phi)}{\dd\Phi} \qty(\Bar{\Psi}_1\Psi_1 + \Bar{\Psi}_2\Psi_2) \mathcal{F} \nonumber\\
 & & \qquad - \frac{2i}{9}
 \frac{\Phi'^2}{A'^2}e^{-A}\qty( \Bar{\Psi}_1'\gamma^r\Psi_1-\Bar{\Psi}_1\gamma^r\Psi_1' - \Bar{\Psi}_2'\gamma^r\Psi_2 +\Bar{\Psi}_2\gamma^r\Psi_2' )\mathcal{F}  \\ \nonumber
 & & \qquad + \frac{i}{2}e^{-A}\partial_{\mu} \qty( \Bar{\Psi}_1'\gamma^\mu\Psi_1-\Bar{\Psi}_1\gamma^\mu\Psi_1' + \Bar{\Psi}_2'\gamma^\mu\Psi_2-\Bar{\Psi}_2\gamma^\mu\Psi_2' ) \widetilde{\mathcal{F}}\\\nonumber
 & & \qquad + \frac{i}{2}e^{-A} \qty( \pd^\mu\Bar{\Psi}_1\gamma^\nu\Psi_1-\Bar{\Psi}_1\gamma^\mu\pd^\nu\Psi_1 + \pd^\mu\Bar{\Psi}_2\gamma^\nu\Psi_2-\Bar{\Psi}_2\gamma^\mu\pd^\nu\Psi_2)h_{\mu\nu}^{TT}  \bigg] \ .
\eea
First, consider the terms on the last line as an explicit example. The terms involving $\Psi_1$ are 
\begin{equation}
\label{eq:contribution1}
    \begin{split}
       & i(\pd^\mu \Psib_1 \gamma^\nu \Psi_1 - \Psib_1 \gamma^\mu \pd^\nu \Psi_1) \\
       & \qquad = p_2^\mu \qty(\psi_L \ub_L(p_2) + \psi_R \ub_R(p_2))   \gamma^\nu \qty(\psi_L u_L(p_1) + \psi_R u_R(p_1)) e^{i(p_2 - p_1)\cdot x} e^{\Phi - 4 A} \\
        &\qquad \quad + \qty(\psi_L \ub_L(p_2) + \psi_R \ub_R(p_2)) \gamma^\mu p_1^\nu  \qty(\psi_L u_L(p_1) + \psi_R u_R(p_1)) e^{i(p_2 - p_1)\cdot x} e^{\Phi - 4 A}\\
        &\qquad = p_2^\mu \qty[\psi_L^2 \ub_L(p_2) \gamma^\nu u_L(p_1) + \psi_R^2 \ub_L(p_2) \frac{\slashed{p_2}}{\mN} \gamma^\nu \frac{\slashed{p_1}}{\mN} u_L(p_1)] e^{ik\cdot x}e^{\Phi - 4 A} \\
        &\qquad \quad + p_1^\nu \qty[\psi_L^2 \ub_L(p_2) \gamma^\mu u_L(p_1) + \psi_R^2 \ub_L(p_2) \frac{\slashed{p_2}}{\mN} \gamma^\mu \frac{\slashed{p_1}}{\mN} u_L(p_1)] e^{ik\cdot x}e^{\Phi - 4 A} \ ,
    \end{split}
\end{equation}
where $p_1$ ($p_2$) is the momentum of the incoming (outgoing) proton and recall that $k=p_2-p_1$ is the exchanged momentum.

Now, the result for $\Psi_2$ follows by changing the handedness of both bound modes. 
Doing this explicitly gives \begin{equation}
\label{eq:contribution2}
    \begin{split}
       & i(\pd^\mu \Psib_2 \gamma^\nu \Psi_2 - \Psib_2 \gamma^\mu \pd^\nu \Psi_2) \\
        &\qquad = p_2^\mu \qty[\psi_R^2 \ub_R(p_2) \frac{\slashed{p_2}}{\mN} \gamma^\nu \frac{\slashed{p_1}}{\mN}  u_R(p_1) + \psi_L^2 \ub_R(p_2) \gamma^\nu u_R(p_1)] e^{ik\cdot x} e^{\Phi - 4 A}\\
        & \qquad \quad +p_1^\nu \qty[\psi_R^2 \ub_R(p_2) \frac{\slashed{p_2}}{\mN} \gamma^\mu \frac{\slashed{p_1}}{\mN}  u_R(p_1) + \psi_L^2 \ub_R(p_2) \gamma^\mu u_R(p_1)] e^{ik\cdot x} e^{\Phi - 4 A}\ .
    \end{split}
\end{equation}
Contributions \eqref{eq:contribution1} and \eqref{eq:contribution2} may be combined using $u_R(p) + u_L(p) = u(p)$. 
We obtain
\begin{equation}
\label{eq:hmunuTTterm}
\begin{split}
    &i \qty( \pd^\mu\Bar{\Psi}_1\gamma^\nu\Psi_1-\Bar{\Psi}_1\gamma^\mu\pd^\nu\Psi_1 + \pd^\mu\Bar{\Psi}_2\gamma^\nu\Psi_2-\Bar{\Psi}_2\gamma^\mu\pd^\nu\Psi_2) \\
    &\qquad =(\psi_R^2 + \psi_L^2) \ub(p_2) (p_2^\mu \gamma^\nu + p_1^\nu \gamma^\mu) u(p_1)  e^{ik\cdot x} e^{\Phi - 4 A}\ ,
\end{split}
\end{equation}
where we also used $\slashed{p}u(p) = \mN u(p)$. In addition, we may write 
\begin{equation}
\label{eq:gammaidentity}
    p^\mu \gamma^\nu = \frac{p^{\{ \mu}\gamma^{\nu \}}}{2}- \frac{1}{4}\eta^\mn p_\alpha \gamma^\alpha \ .
\end{equation}
We contract \eqref{eq:hmunuTTterm} with $h_\mn^{TT}$, and by transverse-tracelessness, the second part of \eqref{eq:gammaidentity} does not contribute. Hence we find
\begin{equation}
\label{eq:term4}
\begin{split}
     &(\psi_R^2 + \psi_L^2) \ub(p_2) (p_2^\mu \gamma^\nu + p_1^\nu \gamma^\mu) u(p_1)  e^{i(p_2 - p_1)\cdot x} e^{\Phi - 4 A} h_\mn^{TT}     \\
     &\qquad \ \ \ \  = 2(\psi_R^2 + \psi_L^2) \ub(p_2) \frac{P^{\{ \mu}\gamma^{\nu \}}}{2} u(p_1)  e^{ik\cdot x}e^{\Phi - 4 A} h_\mn^{TT} \ .
\end{split}
\end{equation}
We simplify the rest of the terms in $S_F^{(3)}$ in a similar manner. The resulting expressions are 
\begin{equation}
\label{eq:terms13}
    \begin{split}
        \bar \Psi_1 \Psi_1 + \bar \Psi_2 \Psi_2 &= 2 \ub(p_2) \psi_R \psi_L u(p_1)e^{i k\cdot x} e^{\Phi - 4 A}\\
        \Bar{\Psi}_1'\gamma^r\Psi_1-\Bar{\Psi}_1\gamma^r\Psi_1' - \Bar{\Psi}_2'\gamma^r\Psi_2 +\Bar{\Psi}_2\gamma^r\Psi_2' &= 2 \ub(p_2) (\psi_L' \psi_R - \psi_L \psi_R') u(p_1)e^{i k \cdot x}e^{\Phi - 4 A}\\
        \Bar{\Psi}_1'\gamma^\mu\Psi_1-\Bar{\Psi}_1\gamma^\mu\Psi_1' + \Bar{\Psi}_2'\gamma^\mu\Psi_2-\Bar{\Psi}_2\gamma^\mu\Psi_2' &= 0 \ .
    \end{split}
\end{equation}
Substituting the results \eqref{eq:term4} and \eqref{eq:terms13} into $S_F^{(3)}$ yields
\begin{align}
        S_F^{(3)} &= \N\!\int\! \dd^5 x \left[  2 \frac{\Phi'}{A'} e^A \frac{\dd \VN(\Phi)}{\dd\Phi} \ub(p_2) \psi_L \psi_R u(p_1) \F - \frac{4}{9} \qty(\frac{\Phi'}{A'})^2 \ub(p_2) (\psi_L' \psi_R - \psi_L \psi_R') u(p_1) \F \right. \nonumber\\
        &\qquad\qquad\qquad +\left. (\psi_R^2 + \psi_L^2) \ub(p_2) \frac{P^{\{ \mu}\gamma^{\nu \}}}{2} u(p_1)  h_\mn^{TT} \right]e^{ik\cdot x} \ .
\end{align}
Further, inserting here \eqref{eq:Fhmunu} and~\eqref{eq:hmunuTT} we find 
\begin{equation}
\begin{split}
    S_{F\,\mathrm{o-s}}^{(3)} &= \N \int \dd^5 x e^{ik\cdot x}\int \frac{\dd^4p}{(2\pi)^4} e^{-ip\cdot x} \\
    &\quad\times \bigg[ 2 \frac{\Phi'}{A'} e^A \frac{\dd \VN(\Phi)}{\dd\Phi} \ub(p_2) \psi_L \psi_R u(p_1) \hat f_s(p,r) \frac{1}{6}(\eta^\mn - p^\mu p^\nu/p^2)\int d^4 y\, e^{ip\cdot y } \delta \eta_{\mn}(y)  \\
    &\quad -\frac{4}{9} \qty(\frac{\Phi'}{A'})^2 \ub(p_2) (\psi_L' \psi_R - \psi_L \psi_R') u(p_1) \hat f_s(p,r) \frac{1}{6}(\eta^\mn - p^\mu p^\nu/p^2)\int d^4 y\, e^{ip\cdot y } \delta \eta_{\mn}(y) \\
    &\quad+ (\psi_R^2 + \psi_L^2) \ub(p_2) \frac{P^{\{ \mu}\gamma^{\nu \}}}{2} u(p_1) \hat{h}_s^{TT}(p,r) \left(- \frac{1}{2}\hat\epsilon_\mn^{TT,\ab}(p)\right) \int d^4 y\, e^{ip\cdot y } \delta \eta_{\ab}(y)\bigg]  \ .
\end{split}    
\end{equation}

%%%%%%%%%%%%%%%%%%%%%%%%%%%%%%%%%%%%%%%%%%%%%%%%%%
\subsection{Cancelation of the divergence}
\label{subapp:cancelationofthedivergence}
%%%%%%%%%%%%%%%%%%%%%%%%%%%%%%%%%%%%%%%%%%%%%%%%%%

In this Appendix we show analytically that the possible pole in \eqref{eq:finalDfactor} cancels. 

As $t \to 0$, $\hat{h}(0,r) = \hat{f}(0,r) = 1$. Since $\int \dd r\, \psi_R^2 = \int \dd r\, \psi_L^2 = 1$, the condition for the cancelation of the divergence becomes
\begin{equation}
    -\mN = \int \dd r \qty(\qty(\frac{\Phi'}{A'}) e^A \frac{\dd \VN(\Phi)}{\dd\Phi} \psi_R \psi_L  - \frac{2}{9}\qty(\frac{\Phi'}{A'})^2\qty(\psi_R \psi_L' - \psi_R' \psi_L)) \ .
\end{equation}
To write this differently, we use the Einstein equation given relation for the second term,
\begin{equation}
    \frac{4}{9}(\Phi')^2 = (A')^2 - A'' \ ,
\end{equation}
and for the first term write
\begin{equation}
    \qty(\frac{\Phi'}{A'})e^A \dv{\VN(\Phi)}{\Phi} = \frac{(e^A \VN)'}{A'} - e^A \VN \ .
\end{equation}
Using these, we arrive at the condition
\begin{equation}
\label{eq:midresult1}
    -\mN = \int \dd r \qty[\qty(\frac{(e^A \VN)'}{A'} - e^A \VN)\psi_R \psi_L - \frac{1}{2} \qty(1 + \qty(\frac{1}{A'})')(\psi_R \psi_L' - \psi_R' \psi_L)] \ .
\end{equation}
Then, we remove dependence on $\psi_{R,L}'$ using their first order equations of motion,
\begin{equation}
\label{eq:firstorderEoMs}
    \psi_L' = \mN \psi_R - e^A \VN \psi_L \ , \qquad \psi_R' = e^A \VN \psi_R - \mN \psi_L \ ,
\end{equation}
with which the last term becomes
\begin{equation}
    \psi_R \psi_L' - \psi_R' \psi_L = \mN(\psi_R^2 + \psi_L^2) - 2 e^A \VN \psi_R \psi_L \ .
\end{equation}
Inserting this in \eqref{eq:midresult1} and simplifying yields 
\begin{equation}
\label{eq:midresult2}
    -\mN = \int \dd r \qty[\frac{(e^A \VN)'}{A'}\psi_R\psi_L - \frac{\mN}{2}\qty(\psi_R^2 + \psi_L^2) + \frac{1}{2}\qty(\frac{1}{A'})'\qty(2e^A \VN \psi_R \psi_L - \mN (\psi_L^2 + \psi_R^2))] \ .
\end{equation}
The  mass term on the left hand side cancels against the first mass term in the square brackets on the right hand side, so one only has to show that the remaining contributions give a total derivative. We integrate the last term by parts and compute the derivative explicitly again using \eqref{eq:firstorderEoMs}. The result is simply
\begin{equation}
    \qty(2e^A \VN \psi_R \psi_L - \mN (\psi_L^2 + \psi_R^2))' = 2 (e^A \VN)'\psi_R \psi_L \ .
\end{equation}
Inserting this in the integral precisely cancels the remaining term in the total expression.

The integration by parts also produces the boundary term
\begin{equation}
\label{eq:boundaryterm}
    \frac{1}{2}\qty[\frac{1}{A'}\qty(2e^A \VN \psi_R \psi_L - \mN (\psi_L^2 + \psi_R^2))]_{r=0}^{r=\infty} \ ,
\end{equation}
which vanishes at both ends of the integration range. In the IR, $A' \approx -2r$ by~\eqref{eq:IHQCDIRasymptotics}, while $e^A \VN \propto r$ by construction of the potential~\eqref{eq:protonpotential}, so that the Schr\"odinger potentials in~\eqref{eq:fermiondecoupled} grow as $r^2$ and the normalizable modes decay as Gaussians. In the UV, $1/A' \approx -r$ and $e^A \VN/A' \sim \VN(\Phi) \to 0$ by~\eqref{eq:IHQCDUVasymptotics} and~\eqref{eq:protonpotential}, whereas for the normalizable proton mode $\psi_L \approx \mN\, r\, \psi_R$ with $\psi_R \to 0$ \eqref{eq:psiasymptotics}. The UV boundary term thus vanishes as well and the divergence of $\D(t)$ cancels analytically.

%%%%%%%%%%%%%%%%%%%%%%%%%%%%%%%%%%%%%%%%%%%%%%%%%%
%%%%%%%%%%%%%%%%%%%%%%%%%%%%%%%%%%%%%%%%%%%%%%%%%%
\section{Asymptotic solutions}
\label{app:IHQCD}
%%%%%%%%%%%%%%%%%%%%%%%%%%%%%%%%%%%%%%%%%%%%%%%%%%
%%%%%%%%%%%%%%%%%%%%%%%%%%%%%%%%%%%%%%%%%%%%%%%%%%

\subsection{IR asymptotics of the background}

In the IR (large coupling $\lambda$), a generic expansion of the potential, which covers physically motivated choices, is \cite{Gursoy:2007cb,Gursoy:2007er}
\begin{equation}
    V(\lambda) \approx v_0 \lambda^{2Q} \log^P(\lambda) \qty(1 + \sum_{i=1}^{\infty} \frac{v_i}{(\log \lambda)^i}) \ .    
\end{equation}
We adopt a convention where the UV expansion coefficients are denoted by $V_i$ (see below) and IR expansion coefficients are denoted by $v_i$, $i=0,1,2,\ldots$. 
The IR expansions for $A(r)$ and $\Phi(r)$ within V-QCD can be found from \cite{Jarvinen:2011qe}: 

\bea
    A & = &  - \qty(\frac{r-r_0}{R})^\alpha  + A_0 - \frac{1}{2} \frac{P}{1-P} \log\qty(\frac{R}{r-r_0}) + \frac{5}{6} + \frac{P}{4} + \frac{1}{2}P \log \frac{3}{2} + \frac{2V_1}{3PV_0} \nonumber\\
    & & + \frac{-52 P^2 V_0^2 + 4 P^3 V_0^2 + 27 P^4 V_0^2 + 64 V_1^2 - 64 P V_1^2 + 128 P V_0 V_2}{288 P (1+P) V_0^2}\qty(\frac{R}{r-r_0})^\alpha \nonumber\\
    & &+ \ldots \\
    \log \lambda & = & \frac{3}{2}\qty(\frac{r-r_0}{R})^\alpha - \frac{5}{4} - \frac{3P}{8} - \frac{V_1}{P V_0} \nonumber\\
    & & + \frac{-20 P^2 V_0^2 - 40 P^3 V_0^2 + 9 P^4 V_0^2 - 64 V_1^2 + 64 P V_1^2 - 128 P V_0 V_2}{192 P (1+P) V_0^2} \qty(\frac{R}{r-r_0})^\alpha \nonumber\\
    & & + \ldots \ ,
\eea
where the parameters $\alpha$ and $A_0$ are defined as
\begin{equation}
    \alpha = \frac{1}{1-P} \ , \qquad e^{A_0} = \frac{2^P 3^{1-P}}{(1-P)\sqrt{V_0}R} \ ,
\end{equation}
and $R$ is an arbitrary scale we set to unity, $R=1$. In our case, we have $Q = 2/3, \ P = 1/2$. Thus,
\begin{equation}
\label{eq:IHQCDIRasymptoticsgeneral}
    \begin{split}
        A(r) & = - \frac{r^2}{R^2} + \frac{1}{2} \log \frac{r}{R} - \log R - \frac{1}{2} \log v_0 + \frac{5}{4} \log  2 + \frac{3}{4} \log 3  \\
        &\qquad + \frac{23}{24} + \frac{4v_1}{3} +\frac{R^2(-173 + 512 v_1^2 + 1024 v_2)}{3456r^2 } + \mathcal{O}\left(\frac{1}{r^4}\right)  \\
        \Phi(r) & = \frac{3}{2}\frac{r^2}{R^2} - \frac{23}{16} - 2v_1 - \frac{R^2(151+512v_1^2 + 1024v_2)}{2304r^2} + \mathcal{O}\left(\frac{1}{r^4}\right) \ .
    \end{split}
\end{equation}

The updated gravity potential we used was given above in (\ref{eq:IHQCDpotnewmaintext}), and has been adopted from \cite{Jarvinen:2022gcc}. Contributions to the IR expansion stem from the last term in the potential (\ref{eq:IHQCDpotnewmaintext}). The connection of the expansion coefficient $v_0$ to the parameters in this formula is $v_0 = \frac{3 V_{\text{IR}}}{4 \pi^{8/3}}$. 
Inserting the numerical values of the parameters, the expansion coefficients up to fourth order become
\begin{equation}
\label{eq:IRexpansionparameters}
    v_0 \approx 0.2562 \ , \qquad v_1 \approx -1.664 \ , \qquad v_2 \approx -1.384 \ , \qquad v_3 \approx -2.303 \ , \qquad v_4 \approx -4.790 \ .
\end{equation}
Inserting these numerical values in~\eqref{eq:IHQCDIRasymptoticsgeneral} gives the asymptotics 
\begin{equation}
\label{eq:IHQCDIRasymptotics}
\begin{split}
    A(r) &\approx -r^2 + 1.11123 + \frac{1}{4} \log (r^2) - \frac{0.0500579}{r^2} + \mathcal{O}\left(\frac{1}{r^4}\right)  \\
    \Phi(r) &\approx \frac{3}{2}r^2 + 1.89014 - \frac{0.0655382}{r^2} + \mathcal{O}\left(\frac{1}{r^4}\right) \ .
\end{split}
\end{equation}

%%%%%%%%%%%%%%%%%%%%%%%%%%%%%%%%%%%%%%%
\subsection{Near-boundary asymptotics of the background}
%%%%%%%%%%%%%%%%%%%%%%%%%%%%%%%%%%%%%%%

In the UV (weak coupling $\lambda$), we expand the potential as 
\begin{equation}
    V(\lambda) \approx \frac{12}{\ell^2} \qty(1 + \sum_{i=1}^\infty V_i \lambda^i) \ .    
\end{equation}
The background UV expansions are found from \cite{Alho:2012mh} and read
\begin{equation}
\label{eq:IHQCDUVasymptoticsgeneral}
    \begin{split}
        A(r) &=  - \log \qty(\frac{r}{\ell}) + \frac{4}{9 \log(r \Lambda)}  \\
        & \qquad + \frac{\frac{1}{162}\qty(95 - \frac{64 V_2}{V_1^2}) + \frac{1}{81} \log \qty(- \log (r \Lambda))\qty(-23 + \frac{64 V_2}{V_1^2})}{\log^2 (r\Lambda)} + \oford{\frac{1}{\log^3 (r\Lambda)}} \\
        \exp\left[\Phi(r)\right] &=  \- \frac{8}{9\log (r\Lambda) V_1} + \frac{\log \qty(- \log (r\Lambda)) \qty(\frac{46}{81}-\frac{128 V_2}{81 V_1^2})}{\log^2 (r\Lambda)V_1 } + \oford{ \frac{1}{\log^3 (r\Lambda)}} \ .
    \end{split}
\end{equation}

For the potential used in this article, the numerical values of the first four coefficients are explicitly
\begin{equation}
\label{eq:UVexpansionparameters}
    V_1 = 0.4953 \ , \qquad V_2 = 1.220 \times 10^{-2}\ , \qquad V_3 = -4.376 \times 10^{-4}\ , \qquad V_4 = 1.570 \times 10^{-5} \ .
\end{equation}

Inserting these values and setting $\ell=1$ the expansions become 
\begin{equation}
\label{eq:IHQCDUVasymptotics}
    \begin{split}
        A(r) &\approx -\log r + \frac{4}{9 \log(r\Lambda)}+ \frac{0.35078 + 0.18733 \log(-\log(r\Lambda))}{\log^2 (r\Lambda)} \\
        \exp\left[\Phi(r)\right] &\approx  - \frac{21.534}{\log (r\Lambda)} - \frac{9.0762 \log \qty(-\log (r\Lambda))}{\log^2 (r\Lambda)} \ .
    \end{split}
\end{equation}

%%%%%%%%%%%%%%%%%%%%%%%%%%%%%%%%%%%%%%%
\subsection{Near-boundary asymptotics of the fermion wave functions}
\label{app:Aasymptotics}
%%%%%%%%%%%%%%%%%%%%%%%%%%%%%%%%%%%%%%%

In order to analyze UV-normalizability of the fermion action \eqref{eq:protonaction}, we search for asymptotic solutions to the Eq.~\eqref{eq:fermiondecoupled} at weak coupling. 

We first expand the left- and right-handed potentials in ~\eqref{eq:protonpotential} near UV, $r\to 0$. We use the UV expansions of $\Phi(r)$ and $A(r)$ presented in~\eqref{eq:IHQCDUVasymptotics}. Expanding in $r, -\log r$, and $\log(-\log r)$ yields asymptotic formulas 
\begin{equation}
    \VL \sim  \beta \frac{\log(-\log(r))^{1/4}}{r^2 (-\log(r))^{2/3}} \ , \qquad \VR \sim  - \beta \frac{\log(-\log(r))^{1/4}}{r^2 (-\log(r))^{2/3}} \ ,
\end{equation}
where $\beta$ is a constant. Hence, the differential equations (\ref{eq:fermiondecoupled}) in the asymptotic limit $r\to 0$ read 
\be
    \psi''_{L,R}(r) \pm \beta \frac{\log(-\log(r))^{1/4}}{r^2 (-\log(r))^{2/3}} \psi_{L,R}(r) \approx 0 \ ,
\ee
as the mass term is subleading, and plus and minus are for the left- and right-handed modes, respectively.  

We first search for the asymptotics of the VEV part of the solution by inserting an {\emph{Ansatz}} $\psi = r \cdot e^{f(\log(-\log(r)))}$ and solving the resulting leading differential equation for the arbitrary function $f$. This gives the asymptotics for the UV-normalizable left- and right-handed modes as
\begin{equation}
\label{eq:asymptoticsolutions1}
    \psi_{L,R}(r) \sim r \cdot \text{exp} \qty(\mp 3 \beta  \qty((-\log(r))^{1/3}  \log(-\log(r))^{1/4})) \ .
\end{equation}

Likewise, we search for the asymptotics of the source part of the solution by setting the {\emph{Ansatz}} to $\psi = e^{f(\log(-\log(r)))}$. Solving again for the leading behavior gives the non-normalizable solutions
\begin{equation}
\label{eq:asymptoticsolutions2}
    \psi_{L,R}(r) \sim  \text{exp} \qty(\pm 3 \beta  \qty((-\log(r))^{1/3}  \log(-\log(r))^{1/4})) \ .
\end{equation}

In general, the leading order asymptotics therefore read
\bea
    \psi_L & \sim & A_L  e^{3 \beta  (-\log r)^{1/3}  \log(-\log r)^{1/4}} + B_L\, r\, e^{-3 \beta  (-\log r)^{1/3}  \log(-\log r)^{1/4}} \nonumber \\
    \psi_R & \sim & A_R  e^{-3 \beta  (-\log r)^{1/3}  \log(-\log r)^{1/4}} + B_R \,r\, e^{3 \beta  (-\log r)^{1/3}  \log(-\log r)^{1/4}}\ , \label{eq:fullUVasymptotics}
\eea
where $A_{L/R}$ and $B_{L/R}$ are \textit{a priori} free constants for the source and VEV asymptotic solutions, respectively. However, given the relations in Eq.~\eqref{eq:fermioncoupled}, there are only two independent parameters. Relations between constants in (\ref{eq:fullUVasymptotics}) can be explicitly solved by inserting the asymptotic solutions in these equations. We find
\begin{equation}
    A_R = B_L/m_n \ , \qquad B_R = - A_L \cdot m_n \ ,
\end{equation}
where $m_n = \sqrt{p^2}$ is the mass of the nucleon state.
Inserting these relations into the asymptotic formulas, the final expressions read
\begin{equation}
\label{eq:psiasymptotics}
\begin{split}
    \psi_L &\sim A_L \cdot \text{exp} \qty(3 \beta  \qty((-\log(r))^{1/3}  \log(-\log(r))^{1/4})) \\
    &\qquad + B_L r \cdot \text{exp} \qty(-3 \beta  \qty((-\log(r))^{1/3}  \log(-\log(r))^{1/4})) \\
    \psi_R &\sim \frac{B_L}{m_n} \cdot \text{exp} \qty(-3 \beta  \qty((-\log(r))^{1/3}  \log(-\log(r))^{1/4})) \\
    &\qquad  - A_L m_n r \cdot \text{exp} \qty(3 \beta  \qty((-\log(r))^{1/3}  \log(-\log(r))^{1/4})) \ . \\
\end{split}
\end{equation}

\bibliographystyle{JHEP}
\bibliography{biblio.bib}

\end{document}